\documentclass[a4]{article}
\usepackage{a4}
\usepackage{amsfonts}
\usepackage{amssymb}
\usepackage{lscape}
\usepackage{cite}
\usepackage{epsfig}
\usepackage{multirow}
\usepackage[all]{xy}
\usepackage{amsmath}
\usepackage{graphicx}
\usepackage{color}
\usepackage{slashed}
\usepackage{youngtab}

\newcommand{\be}{\begin{equation}}
\newcommand{\ee}{\end{equation}}
\newcommand{\ba}{\begin{array}}
\newcommand{\ea}{\end{array}}
\newcommand{\bea}{\begin{eqnarray}}
\newcommand{\eea}{\end{eqnarray}}

\newcommand{\ov}{\overline}
\def\IR{\relax{\rm I\kern-.18em R}}

\def\IP{\relax{\rm I\kern-.18em P}}
\def\inbar{\vrule height1.5ex width.4pt depth0pt}
\def\IC{\relax\,\hbox{$\inbar\kern-.3em{\rm C}$}}

\def\K3{{\bf K3}}

\def\ov{\overline}

\def\n2d{\cN_{V^*}^{\otimes 2}}

\def\IC{\mathbb{C}}

\def\IR{\mathbb{R}}

\def\IP{\mathbb{P}}

\def\cN{{\mathcal N}}

\begin{document}

\baselineskip=14pt

\vspace*{-1.5cm}
\begin{flushright}    % Publication numbers
  {\small
  MPP-2010-87 \\
  LMU-ASC 53/10\\
  NSF-KITP-10-102}
\end{flushright}

\vspace{2cm}
\begin{center}        % Main title
  {\LARGE
FCNC Processes from D-brane Instantons
  }
\end{center}

\vspace{0.75cm}
\begin{center}        % Authors
  Ralph Blumenhagen$^{1,3}$, Andreas Deser$^{2}$, Dieter L\"ust$^{1,2,3}$
\end{center}

\vspace{0.15cm}
\begin{center}        % Institutes 
  \emph{$^{1}$ Max-Planck-Institut f\"ur Physik, F\"ohringer Ring 6, \\ 
               80805 M\"unchen, Germany}
               \\[0.15cm] 
  \emph{$^{2}$ Arnold Sommerfeld Center for Theoretical Physics,\\ 
               LMU, Theresienstr.~37, 80333 M\"unchen, Germany}
   						 \\[0.15cm]
   \emph{$^3$ Kavli Institute for Theoretical Physics, Kohn Hall, \\ UCSB, Santa Barbara, CA 93106, USA }
    
\end{center} 

\vspace{2cm}

%%%%%%%%%%%%%%%%%%%%%%%%%%%%%%%%%%%%%%%%%%%%%%%
%%%%%%%%%%%%%%%%%%%%%%%%%%%%%%%%%%%%%%%%%%%%%%%
%%%%%%%%%%%%%%%%%%%%%%%%%%%%%%%%%%%%%%%%%%%%%%%
%%%%%%%%%%%%%%%%%%%%%%%%%%%%%%%%%%%%%%%%%%%%%%%
%%%%%%%%%%%%%%%%%%%%%%%%%%%%%%%%%%%%%%%%%%%%%%%
%%%%%%%%%%%%%%%%%%%%%%%%%%%%%%%%%%%%%%%%%%%%%%%
%%%%%%%%%%%%%%%%%%%%%%%%%%%%%%%%%%%%%%%%%%%%%%%
%%%%%%%%%%%%%%%%%%%%%%%%%%%%%%%%%%%%%%%%%%%%%%%

\begin{abstract}
Low string scale models might be tested at the LHC directly
by their Regge resonances. For such models it is important
to investigate the constraints  of  Standard Model precision measurements
on the string scale. It is shown  that highly suppressed 
FCNC processes like $K^0$-$\overline K^0$ oscillations
or leptonic decays of the $D^0$-meson provide  non-negligible  
lower bounds  on both the perturbatively and surprisingly 
also non-perturbatively induced 
string theory couplings. We present both the D-brane instanton
formalism to compute such amplitudes and discuss various
possible scenarios and their  constraints on the string scale
for (softly broken) supersymmetric intersecting D-brane models. 
\end{abstract}

\clearpage

\newpage

%%%%%%%%%%%%%%%%%%%%%%%%%%%%%%%%%%%%%%%%%%%%%%%
%%%%%%%%%%%%%%%%%%%%%%%%%%%%%%%%%%%%%%%%%%%%%%%
%%%%%%%%%%%%%%%%%%%%%%%%%%%%%%%%%%%%%%%%%%%%%%%
%%%%%%%%%%%                 %%%%%%%%%%%%%%%%%%%
%%%%%%%%%%%  DOCUMENT BODY  %%%%%%%%%%%%%%%%%%%
%%%%%%%%%%%                 %%%%%%%%%%%%%%%%%%%
%%%%%%%%%%%%%%%%%%%%%%%%%%%%%%%%%%%%%%%%%%%%%%%
%%%%%%%%%%%%%%%%%%%%%%%%%%%%%%%%%%%%%%%%%%%%%%%
%%%%%%%%%%%%%%%%%%%%%%%%%%%%%%%%%%%%%%%%%%%%%%%

\tableofcontents

\section{Introduction}

In the study of string compactifications, 
Type II orientifolds  with (intersecting) D-branes provide a 
promising avenue to
derive the Standard Model (SM)  of elementary particles from strings (for a review see
\cite{Blumenhagen:2006ci}). Here all light particles of the SM arise
from open strings that live on some local stacks of D-branes, which completely fill the \mbox{uncompactified},
four-dimensional space-time and at the same time wrap some of the cycles of the internal compact space.
Very often, the energy scale, associated with the typical extension of the string and  hence called
 the string scale $M_s\sim 1/\sqrt{\alpha'}$, is assumed to be near the Planck mass $M_P$ or
the Grand Unification (GUT) scale. In this case, direct evidence for string theory at low energies
is hard to access.
But  unlike heterotic string compactifications, the D-brane realization of the SM does not make any a priori prediction
about the string scale $M_s$;  in fact, using as basic experimental input the strengths of the four-dimensional gravitational
and gauge interactions, the string scale can be anywhere between  1 TeV and $M_P\simeq 10^{19}$ GeV.

In case the string scale is really around
the corner at the TeV scale, D-brane compactifications offer exciting possibilities for direct collider signatures of string theory.
The most clear signal would be the direct production of excited string recurrences, most notably the 
first open string excited quark $q^*$  and gluon $g^*$
resonances with masses $M_{g^*,q^*}=M_s$. These particles  can be directly produced at a hadron collider due to the collision of quarks and gluons.
As it was shown in \cite{Lust:2008qc,Lust:2009pz}, the
$n$-point tree level string scattering processes of $n$ gluons or of
two quarks and $n-2$ gluons are completely universal, as they only involve
the production or the exchange of Regge states, but are not contaminated by heavy KK (or winding) states, which would describe
the effects of the internal geometry. Therefore the production cross sections and the decay widths of these Regge states can be computed without any reference
to a specific orientifold compactification
\cite{Anchordoqui:2008hi,Anchordoqui:2008di,Anchordoqui:2009mm}.
Analyzing in particular the $2\rightarrow 2$ parton scattering Veneziano like scattering amplitudes involving four gluons or two gluons and two quarks, the non-appearance of direct s-channel Regge production at the Tevatron sets the following absolute and model
independent  lower limit on the string scale: 
\begin{equation} M_s^{\rm Tev.}\geq~ 1 ~{\rm TeV}\, .
\end{equation}
The LHC  will either discover these colored Regge states in dijet events, or in case
of their non-appearance the model independent limits from direct string production will be eventually  pushed up at the LHC to orders of 
\begin{equation}
M_s^{\rm LHC}\geq{\cal O}(5-10~ {\rm TeV})\, .
\end{equation}
Besides direct string production, the  $n$-point amplitudes with at most two
quark fields also lead to new  universal contact interactions inducing small deviations from
processes that are measured within the SM with high precision. However, due to the structure of the Veneziano amplitudes, the lowest deviation from
the SM due to the universal 4-point amplitudes is given in terms of dimension eight operators like 
$M_s^{-4}F^4$ or $M_s^{-4}F^2(\bar\psi\partial \psi)$ \cite{Cullen:2000ef,Lust:2008qc}.
Therefore the bounds on the string scale from the universal contact interactions are rather mild:
\begin{equation}
M_s^{dim=8}\geq{\cal O}(0.7~ {\rm TeV})\, .
\end{equation}

Additional bounds on the string mass scale are expected to arise from $n$-point fermion scattering processes  with $n\geq4$.
These processes are not anymore universal but rather model
dependent, as they also involve the exchange of KK (and winding) particles, whose details depend
on the internal compact space. %Moreover, the $s$-channel excitation of Regge resonances and KK particles is absent in these 4-fermion
%amplitudes. 
However, quark-antiquark scattering is suppressed at the LHC due to the smallness of the quark/antiquark parton distribution functions
compared to the gluons.  Therefore the model dependent contribution to the direct production at the LHC due to higher order fermion scattering processes is
small, and will basically not alter the direct LHC bounds on $M_s$ discussed above. On the other hand, new contact interactions from 4-fermion (or higher fermion) operators will contain more information on $M_s$ and the masses of the KK particles.
The associated effective operators have dimension six and are of the form
$M_s^{-2}\bar\psi\psi\bar\psi\psi$,  and  considerations of SM high precision tests will imply generically
\begin{equation}\label{dim6}
M_s^{dim=6}\geq{\cal O}(1-1000~ {\rm TeV})\, ,
\end{equation}
where the precise bound will depend on the given model.

In particular, ref.\cite{Abel:2003fk}
 investigated the string tree-level contribution of the 4-fermion, dimension six operators in intersecting D-brane models
to flavor changing neutral currents (FCNC's)  such as the famous $\Delta S=2$, $K^0\leftrightarrow\bar K^0$ mixing.
Indeed they derived strong bounds on the string scale that are even of order  100 TeV.
However, as emphasized already, the 4-fermion scattering processes are very model dependent. It can even happen
that certain 4-fermion processes are perturbatively completely forbidden due to the conservation of "global" $U(1)$ symmetries.
These $U(1)$'s arise as the Abelian parts of the $U(N)$ gauge symmetries on the D-brane world volumes, and they are
generically broken by the Green-Schwarz mechanism at the string scale. Nevertheless the associated $U(1)$ charges
of the matter fields can be used as book-keeping devices and must be conserved in string perturbation theory, 
and hence dangerous processes, such as proton decay or FCNC's, can be
absent due these "global" $U(1)$ symmetries. This is analogous to the absence of certain Yukawa couplings or
neutrino Majorana masses at the perturbative level. 

In \cite{Blumenhagen:2006xt,Ibanez:2006da} it was shown that perturbatively forbidden Majorana mass terms or other forbidden terms like
Yukawa couplings \cite{Blumenhagen:2007zk} in orientifold compactifications 
can be nevertheless created by string 
instantons (see also \cite{Florea:2006si} and \cite{Blumenhagen:2009qh} for a recent review).  These stringy instantons correspond to Euclidean Dp-branes that are 
points in Euclidean four-dimensional space-time and are wrapped around internal $(p+1)$-dimensional cycles.
The associated instanton action breaks the above mentioned $U(1)$ symmetries to  discrete subgroups and hence can induce
perturbatively forbidden mass terms and other rare processes. 

 It is the main scope of this paper to compute FCNC-processes in
D-brane models from Euclidean D-instantons.  Here, we will constrain ourselves to those FCNC-processes that are perturbatively forbidden
and can only occur due to non-perturbative D-instantons. In particular we apply the general formalism to the 
$\Delta S=2$, $K^0\leftrightarrow\bar K^0$ mixing and to the leptonic decay process $K_L ^0  \rightarrow e \mu$, $D^0 \rightarrow e\mu$
which are both perturbatively forbidden in some of the D-brane realizations of the SM.
As we will show, the same non-perturbative instantons that contribute to these
processes are also responsible for some
of the forbidden Yukawa couplings in these models. As in \cite{Anastasopoulos:2009mr,Cvetic:2009yh}, we can 
therefore trade the a priori unknown  D-instanton actions against the
measured Yukawa couplings, and in this way we can derive new bounds on the string mass scale from non-perturbatively generated
FCNC-processes.

At this point we already mention one important 
technical point in our discussions, which has however also very interesting phenomenological consequences.
Namely, the relevant 4-fermion operators fall into two different classes:

\vskip0.2cm
\noindent (i) Non-holomorphic, dimension six operators of the form
\begin{equation}
{\cal L}_{\rm nhol}
={1\over M^2_{\rm nhol}}(\bar \psi_L\gamma_\mu \psi_L)(\bar \psi_L\gamma_\mu \psi_L)+{h.c}\, ,
\end{equation}
and other operators of similar structure. On general grounds, one expects that
the associated mass scale is of the order of the string scale (or the KK scale), i.e.
$M_{\rm nhol}\simeq M_s$. They are called non-holomorphic operators, as in a 
supersymmetric realization of the SM they correspond to non-holomorphic 
D-term operators.  They might be
perturbatively forbidden and could therefore be induced by instantons
having the right zero mode structure. These could in principle be 
non-rigid $O(1)$ instantons generating higher derivative F-terms,
$U(1)$ instantons having also the $\tau^{\dot\alpha}$ zero modes
or instanton-anti-instanton pairs \cite{Blumenhagen:2007bn,GarciaEtxebarria:2007zv}.
Mainly, since such  instanton effects are  not yet completely understood,
we will not consider them in our analysis.

\vskip0.2cm
\noindent (ii) Holomorphic, dimension six operators of the form
\begin{equation}\label{hol}
{\cal L}_{\rm hol}
={1\over M^2_{\rm hol}}(\bar \psi_L \psi_R)(\bar \psi_L \psi_R)+{h.c}\, .
\end{equation}
They are similar to mass terms, and in a supersymmetric version of the SM  they originate from F-terms with a corresponding dimension five holomorphic superpotential
\begin{equation}\label{w}
{\cal W}_{\rm hol}
={1\over M_P}\Psi_L^4\, ,
\end{equation}
where $\Psi_L$ denotes left-handed superfields.  We will entirely 
concentrate on the D-brane instanton generated,
holomorphic operators eq.(\ref{hol}) resp. on
the holomorphic superpotential ${\cal W}$ eq.(\ref{w}). As we will discuss, the zero mode structure of the D-instanton must satisfy
certain conditions for such a dimension five superpotential to be present.

FCNC's by  holomorphic operators will raise  two more important aspects. 
As well known e.g. from dimension five proton decay, the instanton generated
superpotential eq.(\ref{w}) leads for canonically normalized fields to an  effective action of the form 
${\cal L}_{dim=5}={1\over M_F}\psi^2\phi^2$ and  needs one additional loop to convert the two scalars $\phi$ into their two fermionic superpartners $\psi$.
Therefore also the supersymmetry breaking mass scale $M_{\rm susy}$ will be
relevant for our discussion, i.e.
\begin{equation}
      M^2_{\rm hol}=M_F\, M_{\rm susy}\; .
\end{equation}
We will show that this field theory loop can be included
in the D-brane instanton calculus of \cite{Blumenhagen:2006xt} in a consistent manner.

Second, it should be noted that the mass suppression in the superpotential (\ref{w}) is given in terms of  the Planck mass
$M_P$ and not in terms of the string scale $M_s$. Taking into account the K\"ahler metrics
for the various matter fields $\Psi$ together with the K\"ahler potential, it follows
 that the mass suppression $M_{F}$ of the 
physical dimension five operator  
is generically not given in terms
of the string scale $M_s$.
For instance, for MSSM realizations on shrinkable cycles, one can argue
that the physical Yukawa couplings should not depend on the
transverse overall volume of the space \cite{Conlon:2006tj,Conlon:2007zza}, leading eventually
to the appearance of the  so-called winding mass
scale \cite{Conlon:2009qa,Conlon:2010ji}:
\begin{equation}
M_{w}\sim R M_s\, .
\end{equation}
Here $R$ is the size of transverse dimensions in string units, 
which describe directions inside the compact space
orthogonal to the SM D-brane quiver. 
Typically, in the so-called Swiss cheese geometries $R$ is of the order of
the size of the entire six-dimensional space, ${\cal V}\simeq R^6$. 
Hence for the LARGE  volume
scenario, $R$ is very large as well, leading to a large  suppression
factor of the holomorphic FCNC-processes:
$M_{F}\gg M_s$.
Consequently, one can generically state that in such LARGE volume models,  
these holomorphic processes are not very dangerous even if
the string scale $M_s$ is low. The appearance of the high mass scale $M_{\rm w}$ instead of the low string scale $M_s$ is a kind
of mirage effect, which is also important for the mirage gauge coupling unification in low string scale models \cite{Conlon:2009qa,Conlon:2010ji}.
However, we will leave our discussion more general and also consider
MSSM realizations on  non-shrinkable cycles, where we expect a
mass suppression $M_s\le M_{F} \le M_{w}$ to occur.

The paper has the following structure: In section two we briefly summarize the
most important aspects of the instanton calculus in Type II string theory
\cite{Blumenhagen:2006xt}. We also emphasize the correct normalization of the
contributions to the effective supergravity action by pointing out that
for the LARGE volume scenario the suppression factor of the
physical operators is given by the  winding scale. 
In addition we discuss an extension of the instanton calculus including 1-loop diagrams in order to get phenomenologically interesting processes which contain only standard model fermions as external particles.
We then turn in section three to the investigation of flavour violating
processes in a concrete 5-stack MSSM model, which appeared in the bottom-up
analysis of \cite{Cvetic:2009yh}. Here D-brane instantons are needed to generate perturbatively forbidden Yukawa couplings. To get the right hierarchy of quark and lepton masses, the suppression factor of the instanton is fixed and therefore the strength of every additional induced phenomenology.  After giving a list of possibly interesting holomorphic operators generated by these instantons, we find contributions to  $K_0\leftrightarrow\bar K_0$ mixing and to the $D^0$ and $K^0$-decay modes $D^0 \rightarrow e^- e^+$, $D^0 \rightarrow \mu e$ and $K^0 \rightarrow \mu e$ and use these to infer additional lower bounds on the string scale $M_s$. Let us emphasize that we are not providing a full string theory realization of the flavour structure of the Standard Model, but just assume that this is possible and then make order of magnitude estimates of the stringy induced FCNC processes.

\section{Ep-instantons in Type II string theory}

In this section we  recall a few useful facts about Type IIA/B D-brane instantons and
the instanton induced holomorphic superpotential couplings (see
\cite{Billo:2002hm,Blumenhagen:2006xt, Ibanez:2006da, Argurio:2007vqa, Bianchi:2007wy, Ibanez:2007rs, Akerblom:2007uc, Blumenhagen:2009qh} 
for more details).
Specifically,  we want to look at instantons which wrap 3- respectively 4-dimensional internal cycles and therefore extended objects in the internal space.

%Let us, for concreteness consider type IIA string theory compactified on a
%Calabi-Yau manifold. To reduce the amount of supersymmetry from
%$\mathcal{N}=2$ to $\mathcal{N}=1$ we perform an orientifold projection
%$\Omega \bar{\sigma} (-1)^{F_L}$, where $\Omega$ denotes the world sheet
%parity operator, $\bar{\sigma}$ is an antiholomorphic involution (such as $z_i
%\mapsto \pm \bar{z}_i$, if $z_i$ are local complex coordinates of the internal
%manifold) and $F_L$ denotes the number operator of left moving fermions. We
%therefore get $O6$-planes wrapping a (special Lagrangian) 3-cycle in the
%internal manifold, whose RR-tadpole is cancelled by additional
%$D6$-branes wrapping in general different 3-cycles of the manifold. 

Let us consider Type II string theory compactified on a Calabi-Yau manifold. To reduce the amount of supersymmetry from $\mathcal{N}=2$ to $\mathcal{N}=1$ we perform an orientifold projection which leads to $O6$-planes wrapping special Lagrangian 3-cycles of the internal manifold in the Type IIA setting and to $O7$- and $O3$-planes wrapping holomorphic 4-cycles or being localized at points in the Type IIB case respectively.
The RR-tadpoles are then canceled by additional $D6$- ($D7$)-branes wrapping in general different 3- (4-)cycles of the internal space in the case of Type IIA(B) string theory.

In general, a stack of such D-branes carry a $U(N)$ Chan-Paton gauge group, 
whose diagonal Abelian subgroup $U(1)\subset U(N)$ is generically anomalous.
This anomaly is canceled via the Green-Schwarz mechanism, which 
leads to a gauging of an axionic shift symmetry involving
the dimensional reduction of the RR-forms $C_3$ and $C_4$. This renders
the Abelian gauge boson massive and degrades it from a local
to a global symmetry. All perturbative string couplings satisfy
this $U(1)$ selection rule, but at the non-perturbative it
is broken by E2/E3-brane instantons, whose instanton actions
contain the Chern-Simons  couplings to $C_3$ and $C4$ 
\begin{equation}
\label{actinst}
S_{E_2} = e^{-\phi} \int_{\Gamma_3} \Omega_3 + i \int_{\Gamma_3} C_3\; ,\qquad
S_{E_3} = e^{-\phi} \int_{\Gamma_4} J \wedge J + i \int_{\Gamma_4} C_4\, ,
\end{equation} 
%\begin{equation}
%\label{actinst2}
%S_{E_3} = e^{-\phi} \int_{\Gamma_4} J \wedge J + i \int_{\Gamma_4} C_4\, ,
%\end{equation}
and therefore, due
to the gauging, also transforms non-trivially under the 
global $U(1)$.
The transformation of the
instanton action under such anomalous $U(1)$ precisely
reflects the $U(1)$ charges of the instanton zero modes \cite{Blumenhagen:2006xt}.
Let us also recall a few facts about such zero modes.

\subsection{Zero modes}

Since the quantum fluctuations of an Ep-instanton are described
by open strings ending on the Ep-brane, one can use open
string techniques to determine them.
For a single instanton sector, one has to distinguish two
kinds of zero modes, those from open strings with both end-points
on the Ep-brane and those from open strings stretched
between the instantonic brane and the D6/D7-branes in the background.
  
Let us first discuss the Ep-Ep sector. Due to being point-like in 
four dimensions, the instanton breaks translational
symmetry in these directions and therefore we get four massless
Goldstone-bosons $x^{\mu}$.  In addition the instanton
will break some of the supersymmetries, which will
lead to Goldstino zero modes. For an instanton wrapping
a (p+1)-cycle not invariant under the orientifold projection
it will break the bulk ${\cal N}=2$ supersymmetry with eight
supercharges to  ${\cal N}=1$. This will result in
four Goldstino zero modes, usually denoted as
$\theta^{\alpha}$ and  $\bar{\tau}^{\dot{\alpha}}$.
For a contribution to an F-term in the four-dimensional
effective action, the instanton is only allowed to  have the two
$\theta^{\alpha}$ zero modes, which can occur if the instanton
wraps a cycle who local bulk supersymmetry is already reduced
to ${\cal N}=1$. Thus it is either in an orientifold invariant
position or on top of one of the $D$-branes. In the first case,
one can show that the orientifold projection must anti-symmetrize
along the world-volume of the instanton, i.e. the instanton must be
a so-called $O(1)$ instanton.  In addition one gets both bosonic
and fermionic zero modes from transverse deformation
of the Ep-brane. 
The multiplicity of these zero modes is counted, for instance for D6-branes,
by the first  Betti-number $b^1_\pm(\Gamma_3)$ of the wrapped cycle, 
and an instanton without these zero modes 
is called \emph{rigid}.

Second, there can be zero modes from open strings between
the Ep-brane and the D-branes. Since in this case
one has four Neumann-Dirichlet boundary conditions, 
the NS-sector zero mode energy is shifted by one-half and
generically one only gets fermionic zero modes.
These are located on the point-like intersection between
the Ep and the D(4+p)-branes and for single chiral intersection
the GSO projection leaves just one  single Grassmannian degree of
freedom.  These zero modes are also called \emph{charged} zero modes 
because they transform with respect to the fundamental representation 
of the gauge group living on the stack of D-branes. 
Since it will be important for the following,
in table \ref{0content} we list
the multiplicities and representations of these charged zero modes. We use the notation $I_{a-b} =I_{a-b}^+ - I_{a-b} ^-$ to denote the topological intersection number in Type IIA and the chiral index in Type IIB \cite{Blumenhagen:2009qh}.

\begin{table}[h]
\centering
\begin{tabular}{|c|c|c|}
\hline
Zero mode & Representation & Multiplicity \\ \hline
$\lambda_a := \lambda_{Ep-a}$ & $(-\mathbf{1}_{Ep}, \tiny \yng(1)_a) $& $I_{Ep-a} ^+$ \\
$\bar{\lambda}_a := \lambda_{a-Ep}$ &$ (\mathbf{1}_{Ep}, \overline{\tiny {\yng(1)}}_a)$ & $I_{Ep-a}^- $\\
$\lambda_a ' := \lambda_{Ep'-a}$ & $(\mathbf{1}_{Ep},\tiny \yng(1) _a)$ & $I_{Ep'-a} ^+ $\\
$\bar{\lambda}_a ' := \lambda_{a-Ep'}$ & $(-\mathbf{1}_{Ep} ,\overline{ \tiny {\yng(1)}} _a)$ & $I_{Ep'-a} ^-$ \\ \hline
\end{tabular}
\caption{Zero mode content of an instantonic Ep-brane intersecting with a stack $a$ of D(4+p)-branes}
\label{0content}
\end{table}

\subsection{Instanton contributions to the superpotential}

In the following section we want to give the prescription of how to find
corrections to the superpotential of the four-dimensional effective Type II
supergravity action. For  our phenomenological considerations it is not
necessary to fully 
calculate the CFT-correlators. It suffices to just read off the general
structure of the correction terms and estimate their order of magnitude. In particular, we are not considering possible extra suppressions from world-sheet instantons in Type IIA respectively  reduced wave function overlaps in Type IIB. For a concrete model, these effects and taking all correct normalization factors of "4$\pi$" etc. into account might allow one to loose or gain one or two orders of magnitude.

Let us therefore first start with  some general considerations about
the involved scales. 
Since we are interested in low string scale models, we have to
be very careful with distinguishing the string scale from the
Planck scale. The string and the Planck scale
are related by $M_p=M_s/\sqrt{\cal V}$, where ${\cal V}$ denotes the
volume of the six-dimensional internal manifold in units
of the string length (the string coupling constant is omitted here). 
When expressed in terms of chiral superfields, ${\cal V}$ is not a holomorphic
quantity, which means that in the supergravity action
all higher dimensional operators are suppressed by the
Planck scale. 
Thus, a superpotential term consisting of a product of chiral
superfields $\Phi_i = \phi_i + \sqrt{2} \theta \psi_i + \theta \theta F_i$
is multiplied by the appropriate factor of the Planck  mass
\begin{equation} 
W = M_p ^{-n} \prod_{i=1} ^{n+3} \Phi_i\; .
\end{equation}
After integrating over the superspace variables and changing to
canonically normalized  fields, this leads to the 
following coupling in the effective action
\begin{equation}\label{supergravity}
{\cal L}=\frac{1}{M_p^n}\frac{e^{K/2} }{\sqrt{K_{ii}
    K_{jj} \prod_{k \neq i,j} K_{kk}}}\;   \psi_i\, \psi_j \prod_{k \neq i,j}
\phi_k \; .
\end{equation}
Here $K$ is the K\"ahler potential and $K_{ij}$ denotes 
the matter field metric which for simplicity we
assume to be diagonal.
 
Now we are interested in the effective scale of these couplings, which
apparently is not  directly related to the Planck scale.
For small string scale, the volume is large so that this can
induce substantial effects. To be more precise, we need to know
how the K\"ahler potential and the K\"ahler metrics scale
with the volume. For the latter this is not generally known,
but for matter fields localized on shrinkable D7-branes, the
independence of the physical Yukawa couplings from the overall
volume fixes the scaling as \cite{Conlon:2006tj}
\begin{equation}
K_{ii} \propto \mathcal{V}^{-\frac{2}{3}}\; .
\end{equation}
In addition, the K\"ahler potential contains a term $K=-2 \ln \mathcal{V}$ 
leading to 
\begin{equation}\label{scale}
{\cal L}\simeq M_P ^{-n} \mathcal{V}^{\frac{n}{3}} ~\psi_i \psi_j \prod_{k\neq
  i,j} \phi_k =  \frac{1}{ \bigl(M_s {\cal V}^{1\over 6}\bigr)^n} 
   ~\psi_i \psi_j \prod_{k\neq i,j} \phi_k \;. 
\end{equation}
where we have used the relation between the string and Planck mass.
The scale appearing in the denominator is the so-called
winding scale $M_w$ \cite{Conlon:2006tj}, which can also be expressed as
$M_w=M_s^{2/3} M_p^{1/3}$.  For $M_s\simeq 1\,$TeV, one
finds for the winding scale $M_w=10^5\,$TeV, which therefore leads
to a substantially larger suppression of  F-term induced
couplings than naively expected. 
Note that this argument only holds for D-branes
wrapped on shrinkable cycles leading to quiver type gauge theories.
In general, without more knowledge of the matter fields metrics,
one at best assumes the worst case, i.e. that the suppression
scale is the string scale $M_s$.

\begin{figure}[h]
\centering
	\includegraphics[width=0.80\textwidth]{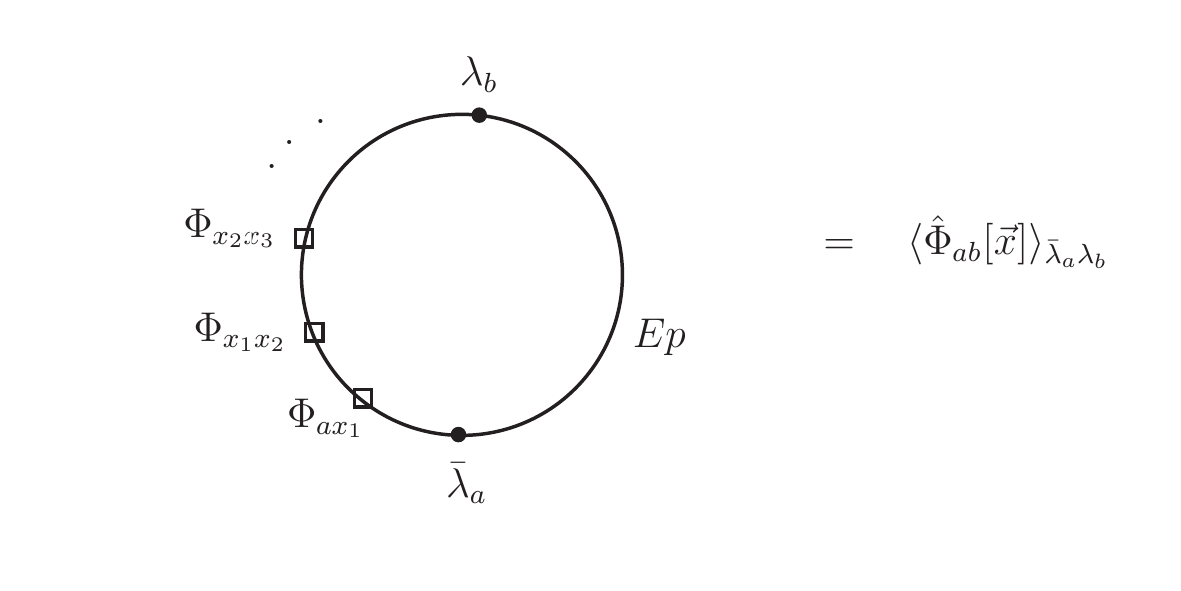}
\caption{\small{Disc diagram with two zero modes $\bar{\lambda}_a$ and $\lambda_b$ which change the boundary from the matter branes to the Ep-instanton brane. In addition matter fields $\Phi_{ij}$ are inserted. Between insertions $\Phi_{ij}$ and $\Phi_{jk}$ the boundary is the matter brane $j$.}}
\label{disc}

\end{figure}

After these general remarks,  let us now turn to the computation of the 
instanton contribution to the
superpotential using the instanton calculus. In the section about the zero
modes of Ep-instantons we have seen that, if the instanton is rigid and lies on
top of an orientifold plane, then it has the right amount of universal zero
modes to have a chance to contribute to the 
superpotential.
In general, if the cycle of the instanton intersects the cycle on which the
matter-brane-stacks are wrapped, there are in addition charged zero modes over
which we have to  integrate.
It has been shown in \cite{Blumenhagen:2007zk} that the contribution
to the superpotential can be extracted from  the following
correlation function in the instanton background
\begin{align}
&\langle \Phi_{a_1 x_{11}} \Phi_{x_{11} x_{12}} \dots \Phi_{x_{1 n_1} b_1} \dots \Phi_{a_M x_{M1}} \Phi_{x_{M1} x_{M2}} \dots \Phi_{x_{M n_M} b_M} \rangle \nonumber \\
&\cong \int d^4 x d^2 \theta \sum_{conf} \prod_a \; \prod_{i=1} ^{I_{Ep D_a} ^+}
  d\lambda _a ^i \; \prod_{i=1} ^{I_{Ep D_a } ^-} d \bar{\lambda} _a ^i  \nonumber \\ 
& \quad e^{-S_{\rm inst}}   e^{Z_{1-loop}} 
\langle \hat{\Phi}_{a_1 b_1} [\vec{x}_1]\rangle_{\lambda_{a_1}
  \bar{\lambda}_{b_1}} \dots \langle \hat{\Phi}_{a_L b_L}
        [\vec{x}_L]\rangle_{\lambda_{a_L} \bar{\lambda}_{b_L}}\; , 
\end{align}
where we have introduced the notation
\begin{equation}
\hat{\Phi}_{a_k b_k} [\vec{x}_k] := \Phi_{a_k x_{k1}} \Phi_{x_{k1} x_{k2}} \Phi_{x_{k2} x_{k3}} \dots \Phi_{x_{k (n_k -1)} x_{k n_k}} \Phi_{x_{k n_k}b_k} 
\end{equation}
and the correlators are pictorially given by disc diagrams as depicted in figure \ref{disc}.

\noindent
The amplitude contains both an exponential of the the classical
instanton action $S_{\rm inst}$ and of the vacuum one-loop partition function
$Z_{\rm 1-loop}$ with at least one boundary on the instanton. This factor
is nothing else than the one-loop determinant of fluctuations
around the instanton background.
In the case of an Ep-instanton, the classical instanton action
is given by eq.\eqref{actinst}.

\vskip 10pt
Summarizing, D-brane instantons with the right number of zero modes
generate contributions to the matter field superpotential
of the following schematic form
\begin{equation}\label{contribution}
W= {1\over M_P^n} \int d^4 x \, d^2 \theta\;  \prod_{i=1} ^{n+3} \Phi_{a_i b_i}\; A(U)\; e^{-S_{Ep}}
\end{equation}
where $A(U)$ is a moduli dependent one-loop determinant.
%and
%$M_w$ denotes the scenario dependent suppression scale, which 
%is not necessarily to be identified with the string scale. 

\subsection{Loop dressing of instanton induced F-terms}

\begin{figure}[h]
\centering
		\includegraphics[width=0.6\textwidth]{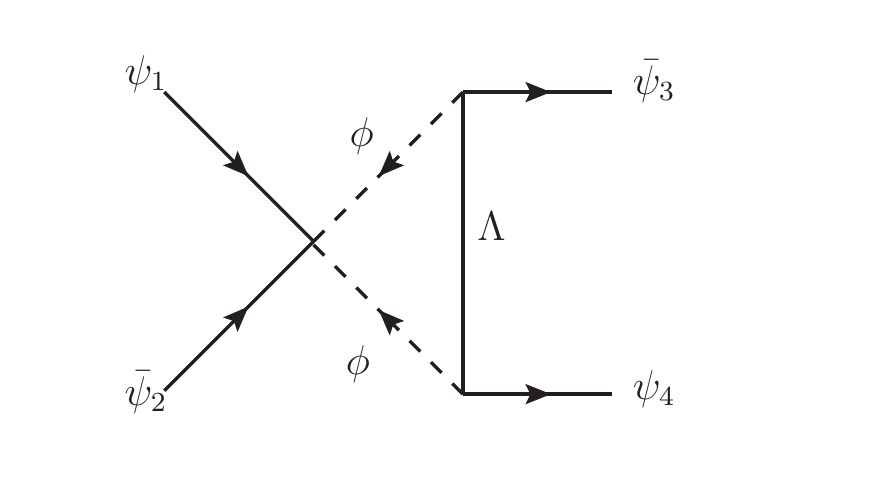}
	\caption{\small{Coupling of the instanton contribution to the MSSM}}
\label{vertexgen}
\end{figure}

It is a general feature of supersymmetric extensions of the
Standard Model that new  higher dimensional F-terms 
do not directly lead to new Standard Model processes,
as only two of the external states are fermions and
all the remaining ones are bosonic superpartners. Thus one has
to dress these effective new interactions by additional
loops, which convert the bosonic superpartners into
the Standard Model fermionic matter particles.
This is quite well known for supersymmetric dimension five
proton decay operators and also for the supersymmetric FCNC
operators of main interest in this paper.
The standard field theory diagram of such a loop diagram is 
shown in figure \ref{vertexgen}.
For the external particles to be all fermions, the propagators
in the loop must involve two bosonic and one fermionic superpartner
of Standard Model particles. In the MSSM the two vertices
on the right hand side must clearly satisfy the selection rules
of the MSSM. 

However, in string theory respectively intersecting D-brane
realization of the MSSM, one has these additional global $U(1)$
symmetries, which all perturbative couplings have to satisfy.
Therefore, in a concrete MSSM like D-brane model, not
every MSSM diagram necessarily is present in string theory.
In order to see directly what is possible, it is helpful
to know  the complete  string theory diagram, 
i.e. the diagram which involves both the instanton
induced higher dimensional F-term and the additional
perturbative loop. 

\begin{figure}[h]
\centering
	\includegraphics[width=0.7\textwidth]{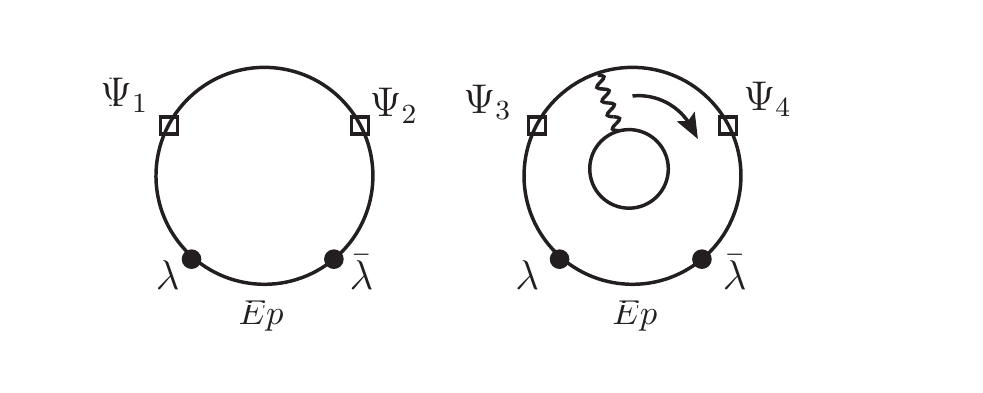}
	\caption{\small{String theoretic realization of the process}}
\label{annulusgen}

\end{figure}

Thus, we are looking for an extension of the instanton
calculus from the last section, which directly involves
the loop. Cutting the above diagram by a vertical line
going through the dimension five vertex, it is
clear what we should do. As the left hand side involves
the external particles and the instanton, it is just 
the usual absorption of charged instanton zero modes
via disc diagrams, as discussed in the last section.
Contrarily, the right hand side of the diagram involves
the external particles and the instanton connected
by a loop. Thus instead of absorbing the fermionic
zero modes by disc diagrams, we do it by a one-loop annulus
diagram as shown in figure \ref{annulusgen}. 
Since we have two fermionic matter zero modes
on the annulus, this diagram does not contribute to the
superpotential, which is of course consistent with the
fact that the entire process is a dimension six operator with
four external fermions.

The internal propagators can then easily be read
off from the  nature of the open strings stretching
between the internal boundary and the various segments
of the external boundary. Since this is a consistent string
diagram, all selection rules are automatically satisfied.
It is clearly not  an easy task to compute the whole
instanton diagrams, but for our purposes it is not
necessary, as we are mainly interested in  the selection rules.

\begin{figure}[ht]
  \centering
  
	\includegraphics[width=0.70\textwidth]{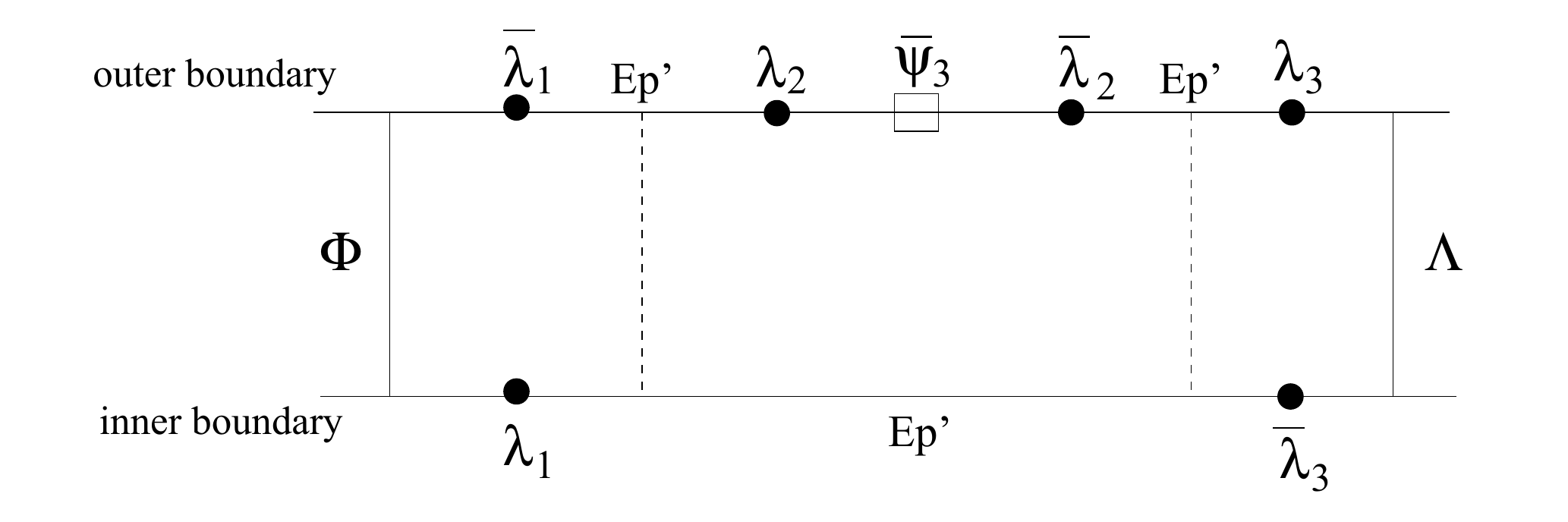}

  \caption{\small non-perturbative Yukawa vertex: the dashed lines indicate
 cuts one can make in order to visualize  the three disc diagrams \eqref{discH96}}
  \label{fig:nploop}
\end{figure}

What we have just discussed is the string theory 
diagram corresponding to a field
theory diagram where a single   vertex is non-perturbatively realized in 
string theory. Clearly, this can be generalized to the case that
more than one field theory vertex is generated by instantons.
In this case the  whole diagram  is a multi-instanton process 
in string theory and one has to integrate over all the appearing
zero modes leading to  a multi-exponential suppression. 
Without developing the whole formalism at this
point, let us discuss an example, which shows the main
features and will be relevant in the following.
Let us assume that the upper right vertex $(\phi,\ov\psi_3,\Lambda)$
in  the field theory diagram in figure \ref{vertexgen} is generated  
non-perturbatively by another Ep-instanton in string theory.
Moreover, we assume this  Yukawa interaction is generated
by an instanton with six matter zero modes, so that in the
instanton calculus one needs three disc diagram
\bea
\label{discH96}
             \langle  \Phi \rangle_{\lambda_1 \ov \lambda_1}, \quad 
               \langle  \Psi_3 \rangle_{\lambda_2 \ov \lambda_2}, \quad 
               \langle  \Lambda \rangle_{\lambda_3 \ov \lambda_3}
\eea
to absorb all  matter zero modes. The question is, how 
all these three couplings can be put into the loop diagram
in figure \ref{annulusgen}.  Clearly, the change can only involve
the closer environment of the $\ov\psi_3$ insertion.         
We propose that indeed the D-brane instanton  calculus
is rich enough to be able to also include such cases
and that the annulus diagram in \ref{annulusgen} is 
changed close to the $\ov\psi_3$ insertion in the way shown in figure
\ref{fig:nploop}.

\noindent
Let us emphasize again that being able to write  such diagrams directly
in an extended instanton calculus in  string theory 
ensure that all global $U(1)$ selection rules are automatically
satisfied. 

We will see that such diagrams appear for instanton induced
higher dimensional F-terms and their one-loop
conversion to observable amplitudes. We will
use these diagrammatic techniques in our
discussion of FCNCs in TeV scale D-brane models.

\section{Flavor-violation in 5-stack quivers}

So far our discussion has been quite general and more on the technical
level. Let us now move forward to discuss the appearance of
perturbative and non-perturbative FCNCs in intersecting
D-brane models. In a bottom-up approach, quasi-realistic models of the MSSM in terms of intersecting branes were studied for example in \cite{Anastasopoulos:2009mr} and \cite{Cvetic:2009yh}.  In order to demonstrate how the techniques described
so far can be utilized, we will concentrate on a very specific example,
namely one of the promising looking five-stack models of 
\cite{Cvetic:2010mm}.

\subsection{The 5-stack quiver and its Yukawa couplings}

Intersecting D-brane compactifications have the potential to realize the particles of the SM at low energies without chiral exotics.
Specifically, open strings that are located on local D-brane quivers give rise to the SM degrees of freedom, where the D-brane quiver 
typically contains three, four, five or even more stacks of D-branes. As a demonstrative example, we will use in the following
the 5-stack quiver, which was also recently
investigated 
in \cite{Cvetic:2010mm}  as a local model
that comes close to the MSSM.
In this model the hypercharge is given by the linear combination
\begin{equation}
U(1)_Y = \frac{1}{6} U(1)_a + \frac{1}{2} U(1)_c + \frac{1}{2} U(1)_d +
\frac{1}{2} U(1)_e \; .
\end{equation}
One choice of matter content is given in table \ref{5stack}.
\begin{table}[h]
\centering 
\caption{Spectrum of the 5-stack quiver}
\label{5stack}
\vskip0.3cm
\begin{tabular}{|c|c|l|c|c|}
\hline
Sector & Matter  & Representation & Multiplicity & Hypercharge \\
\hline
%&&&& \\
$ab$ & $ Q_L ^1 $ & $(\bf{3},\bf{2})_{(1,-1,0,0,0)} $ & $ 1 $ & $\frac{1}{6} $\\
$ab'$ & $Q_L ^{2,3} $ & $ (\bf{3},\bf{2})_{(1,1,0,0,0)} $ & $ 2 $ & $\frac{1}{6} $\\
$ac$ & $D_R $ & $ (\bar{\bf{3}},\bf{1})_{(-1,0,1,0,0)}$ & $3$ & $\frac{1}{3} $ \\
$ac'$ & $U_R ^{1,2} $ & $(\bar{\bf{3}}, \bf{1})_{(-1,0,-1,0,0)}$ & $2 $ & $ -\frac{2}{3} $ \\
$ae'$ & $U_R ^3$ & $(\bar{\bf{3}},\bf{1})_{(-1,0,0,0,-1)}$ & $1$ & $-\frac{2}{3}$ \\
$bc$ & $H_u$ & $(\bf{1},\bf{2})_{(0,-1,1,0,0)} $ & $1$ & $\frac{1}{2} $\\
$bc'$ & $ H_d $ & $(\bf{1},\bf{2})_{(0,-1,-1,0,0)}$ & $1$ & $-\frac{1}{2} $\\
$bd'$ & $L^{1,2} $ & $(\bf{1},\bf{2})_{(0,-1,0,-1,0)}$ & $2 $ & $-\frac{1}{2} $\\
$be$ & $L^3$ & $(\bf{1},\bf{2})_{(0,1,0,0,-1)}$ & $ 1$ & $ -\frac{1}{2}$ \\
$cd'$ & $E_R ^1 $ & $(\bf{1},\bf{1})_{(0,0,1,1,0)}$ & $1$ & $1$\\
$ce'$ & $E_R ^{2,3}$ & $(\bf{1},\bf{1})_{(0,0,1,0,1)}$ & $2$ & $1$\\
\hline
\end{tabular}
\end{table}
\noindent
The quark and lepton Yukawa couplings have the following $U(1)$-charges :
\begin{align}
&Q_L ^1 U_R ^{12} H_u &0,-2,0,0,0 \nonumber \\
&Q_L ^1 U_R ^3 H_u &0,-2, 1, 0 , -1 \nonumber \\
&Q_L ^{2,3} U_R ^{1,2} H_u &0, 0, 0, 0, 0 \nonumber \\
&Q_L ^{2,3} U_R ^3 H_u &0, 0, 1, 0, -1 \nonumber \\
&Q_L ^1 D_R H_d &0, -2, 0, 0, 0, \nonumber \\
&Q_L ^{2,3} D_R H_d &0, 0, 0, 0, 0 \nonumber \\
&L^{1,2} H_d E_R ^1 &0,-2,0,0,0 \nonumber \\
&L^{1,2} H_d E_R ^{2,3} &0,-2,,0,-1,1 \nonumber \\
&L^3 H_d E_R ^1 &0,0,0,1,-1 \nonumber \\
&L^3 H_d E_R ^{2,3} &0,0,0,0,0 
\end{align}
We observe that the  quark Yukawa couplings $Q_L ^{23} U_R ^{23} H_u$ and $Q_L
^{23}D_R H_d$ are realized perturbatively and are therefore not suppressed by
the exponential factor coming from D-instantons. Thus $Q_L ^{23}$ should be
identified with  the two heavier families. Similar arguments hold for the perturbatively realized lepton-Yukawa coupling $L^3 H_d E_R ^{23}$.

The other  Yukawa couplings have to be realized non-perturbatively by
D-instantons which have the right charges (and thus intersection
structure). These instantons not only generate the missing Yukawa couplings
but also (we have to sum over all possible configurations) give rise to other
phenomenologically important couplings.  In the following section we are going
to investigate the consequences of these additional couplings on processes
which are highly suppressed or even completely forbidden in the standard
model. As usual in a bottom-up approach, we will always assume that
the respective instantons are really present in the model, which for
global intersecting D-brane models provide further severe
constraints.

\subsection{Non-perturbative higher dimensional operators}

We now begin to investigate possible flavour violating effects in MSSM models. First, there are perturbatively allowed contributions to quark flavour mixing, as pointed out in \cite{Abel:2003fk}. They are of the four fermion-type and therefore suppressed by the square of the string mass scale $M_s$:
\begin{equation}
\frac{1}{M_s ^2} \langle \bar{\psi}\psi\bar{\psi}\psi \rangle\; .
\end{equation}
%{\bf ralph: etwas umgeschrieben. bitte checken}
If in an MSSM-like  model the two lightest  families are realized by the same
representation, these four-fermion operators can contribute to flavour
mixing. In the systematic search for realistic MSSM orientifold models done in
\cite{Cvetic:2009yh}, in every model this is indeed the case.  The strongest
bounds come from neutral Kaon mixing and were calculated in
\cite{Abel:2003fk}. It was found that in case  
 the two lightest families are realized by the
same representation, they give a lower bound on the string scale of $M_s >
10^{3-4}$ TeV. Note that for low string scale models this bound is 
already quite  severe and essentially rules out 1TeV scale D-brane models
of this type. Thus for rescuing a TeV scale scenario, we need to assume
that these perturbative couplings are  absent.
However, even then there is the danger that further exponentially
suppressed non-perturbative, i.e. D-brane instanton induced, FCNC
couplings  do not lower the bounds sufficiently enough. 
The evaluation of these bounds is the central question of this
section.

\vskip0.2cm

To illustrate this, consider first the  famous flavour violating process  $K^0\leftrightarrow\bar K^0$ with $\Delta S=2$.
In the SM, this process is induced by the 1-loop box-diagram, which involves the exchange of two $W$-bosons that couple $SU(2)$ doublets.
It corresponds to an effective, dimension six 4-fermion operator of the form
\begin{equation}
{\cal L}_{K^0\leftrightarrow\bar K^0}\sim (\bar s_Ld_L)(\bar s_Ld_L)
\end{equation}
Now, depending on how we define three different quark flavors, there are two ways to realize this 4-fermion operator
by the fields in the table of the five stack model. The first realization is
\begin{equation}
(A):\qquad {\cal L}_{K^0\leftrightarrow\bar K^0}
=(\bar Q_L^{2,3}Q_L^{2,3})(\bar Q_L^{2,3}Q_L^{2,3})\qquad \Delta Q_b=0
\end{equation}
and the second one
\begin{equation}
(B):\qquad {\cal L}_{K^0\leftrightarrow\bar K^0}
=(\bar Q_L^{2,3}Q_L^1)(\bar Q_L^{2,3}Q_L^1)\qquad \Delta Q_b=4\; .
\end{equation}
Apparently, the operator (A) is invariant under all $U(1)$ symmetries
and is therefore present perturbatively. However, if we assign the 
the two heavier families to $Q_L^{2,3}$, this operator does not induce
$K^0\leftrightarrow\bar K^0$ oscillations, but rather 
$B_s^0\leftrightarrow\bar B_s^0$ oscillations. 
As a perturbative 4-point coupling
this dimension six operator is suppressed by $M_s^{-2}$.

The operator (B)  violates $U(1)_b$ of the five-stack model, as it has charge
$\Delta Q_b=4$.  
The reason behind this is that one needs
also the Cabbibo angle, when going from the quark flavor to the quark mass eigenstates. 
 This means that $K_0\leftrightarrow\bar K_0$
implicitly also contains two off diagonal Yukawa couplings. Indeed, some of the entries in the quark Yukawa coupling matrix
 of the five-stack model are perturbatively
allowed, others violate some of the $U(1)$ symmetries. 
Therefore, the $K_0\leftrightarrow\bar K_0$ operator (B) can only be generated
non-perturbatively via  D-instantons. 
%On the other hand, the $K_0\leftrightarrow\bar K_0$ operator (A) does not
%need non-perturbative D-instantons, as it only contains perturbatively
%generated Yukawa couplings.
As we have discussed, the suppression of this non-perturbative operator
depends on how it is precisely generated by instantons and
on the scenario how the MSSM is embedded.

Next we turn to the discussion of the holomorphic flavor violating operators, which arise from a D-instanton induced holomorphic superpotential.
We are looking for a classification of all possible holomorphic chain products of matter fields which are possible insertions in disc amplitudes together with the charged zero modes coming from the intersection of the Yukawa-instantons with the matter branes. Combining them according to the zero mode structure of the instantons, we get the whole set of holomorphic operators (we did the analysis up to mass dimension 5) generated by the instantons. In the following, we list the most interesting results:
\begin{enumerate}
\item  \fbox{$0,-2,0,0,0$}

 If the charges are realized with two zero modes $\lambda_b , \lambda_b$ we get the following interesting dimension 5 operators:
\begin{align}\label{ph}
& Q_L ^1 D_R U_R ^{12} Q_L ^1  & Q_L ^1 U_R ^{12} D_R Q_L ^1 \nonumber \\
&Q_L ^1 U_R ^{12} E_R ^1 L^{12}  & H_d E_R ^{23} U_R ^3 Q_L ^1 \nonumber \\
&L^{12} E_R ^{23} U_R ^{12} Q_L ^1 \; .
\end{align}  
The second way to realize this charge structure is with four zero modes $\lambda_b , \lambda_b , \lambda_c , \bar{\lambda}_c $.  If we combine two discs with $\lambda_b \lambda_b $ and $\lambda_c \bar{\lambda}_c$ inserted, there will be no operators of mass dimension lower than or equal to 5. But if we take one disc with $\lambda_b, \lambda_c$ and one with $\lambda_b, \bar{\lambda}_c$ we get the following operators:
\begin{align}\label{vectorlike}
& Q_L ^1 U_R ^{12} Q_L ^1 D_R  &Q_L ^1 U_R ^{12} L^{12} E_R ^1 \nonumber \\
&H_d Q_L ^1 U_R ^3 E_R ^{23} \; .
\end{align}

\item  \fbox{$0,-2,1,0,-1$}

If we combine $\lambda_b \lambda_b$ with $\bar{\lambda}_c \lambda_e $ there are no operators of dimension smaller or equal to 5. Combining $\lambda_b \bar{\lambda}_c$ and $\lambda_b \lambda_e$ we get the following operators:
\begin{align}
&Q_L ^1 D_R Q_L ^1 U_R ^3 &L^{12} E_R ^1 Q_L ^1 U_R ^3 \nonumber \\
&Q_L ^1 D_R H_u H_d \; .
\end{align}

\item \fbox{$0,0,1,0,-1$}

If we take the direct realization of the charges via $\bar{\lambda}_c \lambda_e$ we get the following operators:
\begin{align}
&D_R Q_L ^1 L ^3 &H_u L^3 \nonumber \\
&E_R ^1 L^{12} Q_L ^{23} U_R ^3 &E_R ^{23} U_R ^3 Q_L ^1 L^3 \; .
\end{align} 
The first line would give dangerous R-parity violating couplings!

\item \fbox{$0,-2,0,-1,1$}

Only the combination of discs with $\lambda_b \lambda_d$ and $\lambda_b \bar{\lambda}_e$ gives operators of dimension smaller or equal to 5. They are:
\begin{align}
&L^{12} Q_L  ^1  U_R ^{12} E_R ^{23} &L^{12} H_d E_R ^{23}
\end{align}

\item \fbox{$0,0,0,1,-1$}

In this case, if we take the direct realization of the charges via $\bar{\lambda}_d \lambda_e$ we get the following operators:
\begin{align}
&E_R ^1 U_R ^{12} Q_L ^1 L^{12}  &E_R ^1 H_d Q_L ^{23} U_R ^3\; .
\end{align}
For the second realization via $\bar{\lambda}_d \lambda_e \lambda_c \bar{\lambda}_c $ there are 3 different possibilities to distribute the zero modes. In the case of $\bar{\lambda}_d \lambda_e$ on the first and $\lambda_c \bar{\lambda}_c $ on the second disc there are no operators of dimension 5 or lower. The combination $\bar{\lambda}_d \lambda_c$ and $\lambda_e \bar{\lambda}_c $ also gives no interesting operators. Finally, the combination $\bar{\lambda}_d \bar{\lambda}_c$ and $\lambda_e \lambda_c$ gives the following operators:
\begin{align}
&E_R ^1 L^3  Q_L ^1 U_R ^{12} &E_R ^1 L_3 H_d \nonumber \\
&E_R ^1 U_R ^3 Q_L ^{23} H_d \; .
\end{align}

\end{enumerate}
\vskip 10pt

\subsection{Quark flavour violation}

\begin{figure}[h]
\centering	\includegraphics[width=0.59\textwidth]{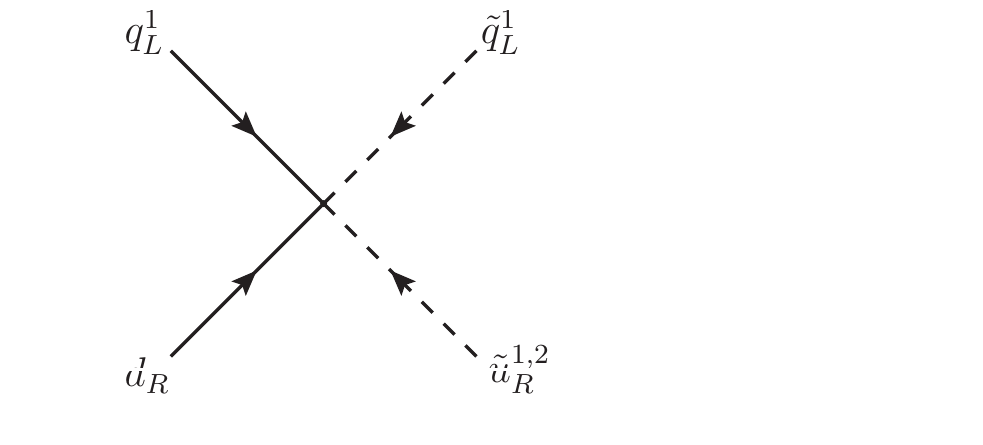}
\caption{\small{Instanton vertex: Solid lines represent fermions and dashed lines represent the bosonic superpartners \newline ~}}
\label{quarkvertex1}
\end{figure}

Let us now focus on the possible contributions to flavour changing processes
in the quark sector. Here we concentrate on neutral meson mixing because this
is very well studied experimentally.
Consider the contribution $Q_L^1\, U_R^{12}\,  Q_L^1\,  D_R$ from eq.(\ref{vectorlike}). Taking the F-term, it contains the following interaction (the tilde again denotes the superpartner to the fermion), as depicted in figure \ref{quarkvertex1}.

\noindent
The strength of this vertex includes the instanton suppression factor and for
dimensional reasons the inverse of the characteristic mass scale $M_F$, i.e.
a scale between the string scale $M_s$ and the winding scale
$M_{w}=M_s^{2/3} M_p^{1/3}$.

Next we discuss how to convert the holomorphic dimension five operator into a dimension six operator that involves four
fermion fields. This can be done by inserting a Yukawa-coupling-vertex
$\bar{Q}_L ^1 \bar{H}_d \bar{D}_R$ at the upper right and a vertex $\bar{Q}_L
^{23} \bar{H}_u \bar{U}_R ^{12}$ on the lower right. Thus  we get as
out-states the fermionic components  of the superfields $\bar{D}_R$ and
$\bar{Q}_L ^{23}$. In addition, to complete the diagram we have to insert a $\mu$-term $\mu H_d H_u$. The first Yukawa coupling and the $\mu$-term are perturbatively forbidden and therefore need an instanton to be generated.
\vskip 10pt

\begin{figure}[h]
\centering
		\includegraphics[width=0.70\textwidth]{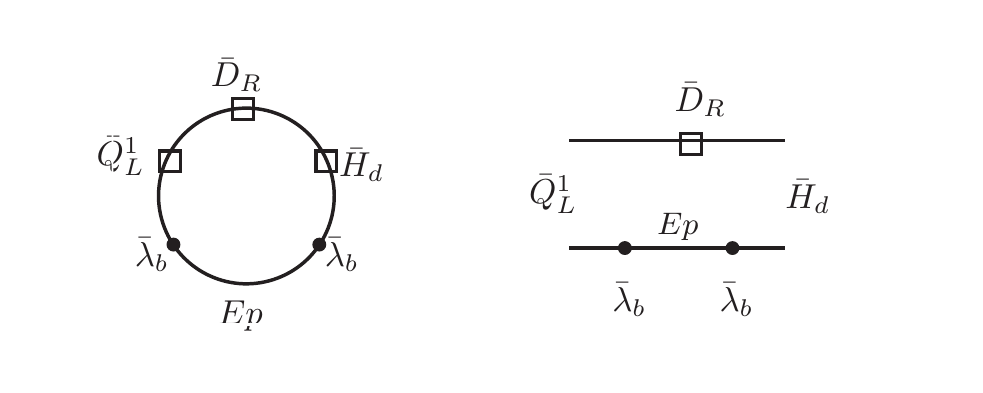}
\caption{\small{The non-perturbative Yukawa coupling and the corresponding piece in an annulus diagram}}
\label{npyuk}
\end{figure}

To realize the non-perturbative Yukawa coupling we use the techniques
presented  in section  2. The easiest possibility is to take the minimal
amount of charged zero modes, i.e. $\lambda_b, \lambda_b$. The disc and the
corresponding piece in the annulus are given in figure \ref{npyuk}.

\begin{figure}[h]
\centering
	\includegraphics[width=0.80\textwidth]{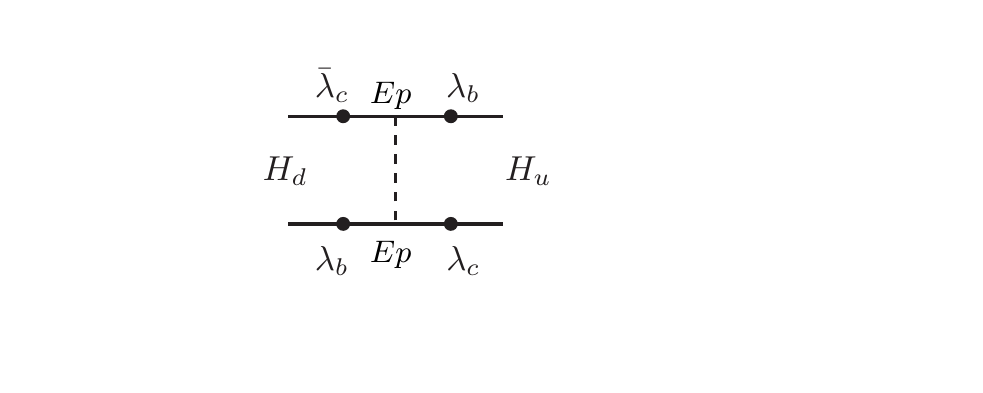}
	\caption{\small{Realization of the $\mu$-term as part of an annulus}}
\label{muterm}
\end{figure}

\noindent
In the same way, we realize the $\mu$-term of the MSSM superpotential as part of an annulus diagram as given in figure \ref{muterm}.

\begin{figure}[h]
\centering
		\includegraphics[width=0.80\textwidth]{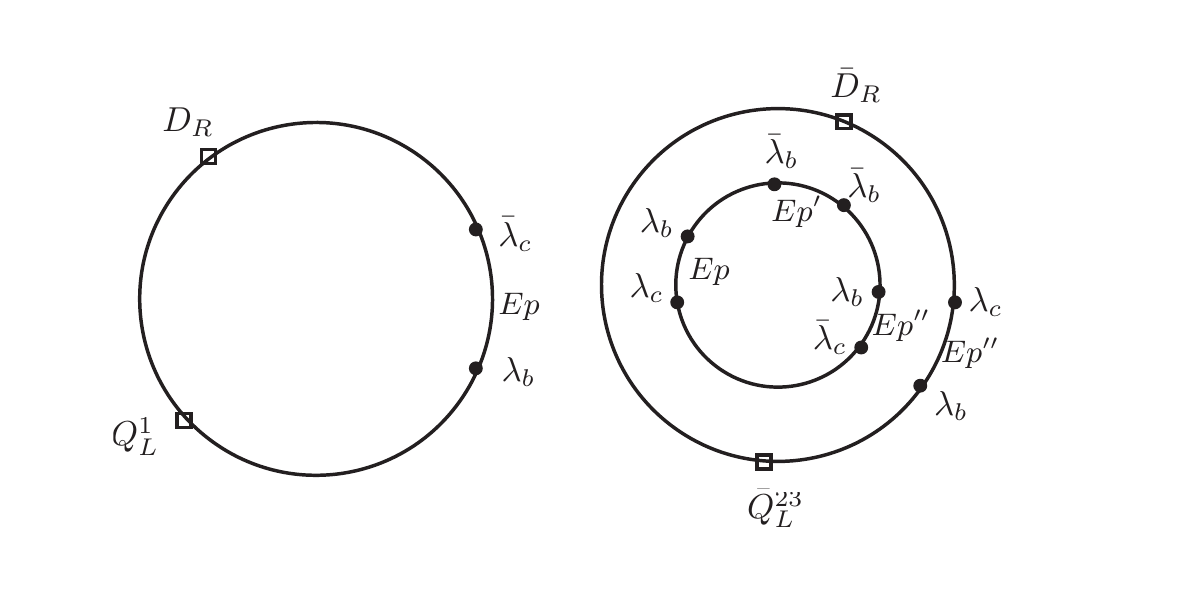}
\caption{\small{3-instanton contribution to neutral Kaon mixing}}
\label{3inst}
\end{figure}

\noindent
To close the annulus diagram, we combine the non-perturbative Yukawa coupling
and the $\mu$-term with the perturbatively realized Yukawa coupling $\bar{Q}_L
^{12} \bar{U}_R ^{12} \bar{H}_u$. In addition, we insert the charged zero
modes $\lambda_b , \lambda_c$ at the inner boundary.
% to couple the annulus diagram to the holomorphic disc (which is required to
% generate the left half of the instanton vertex in figure \ref{quarkvertex1})
% after the zero mode integration. 
As shown in   figure \ref{3inst}, the whole three-instanton process  has only standard model fermions as external states. 
\vskip 50pt

\begin{figure}[h]
\centering
	\includegraphics[width=0.75\textwidth]{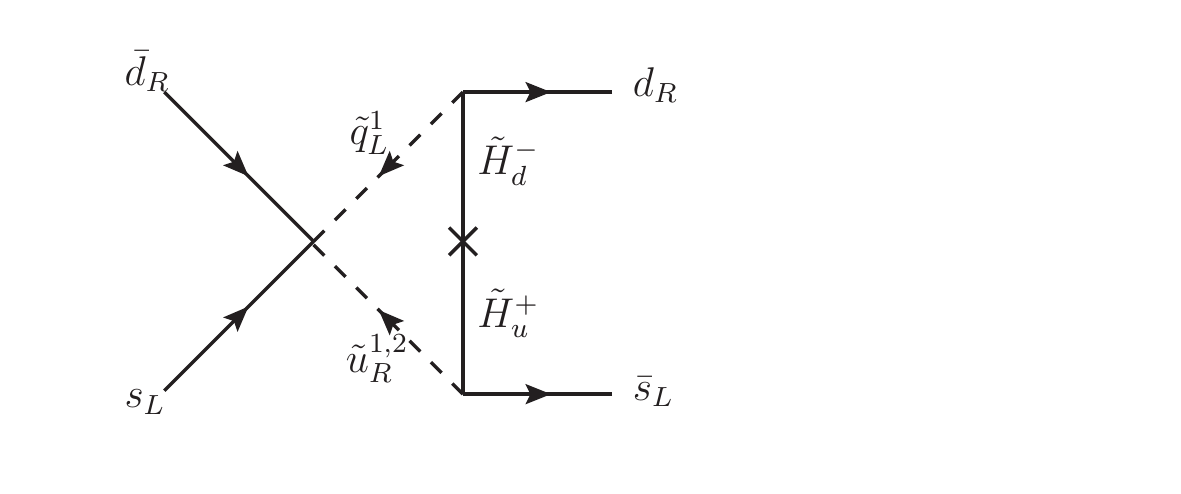}
\caption{\small{Coupling of the instanton contribution to the MSSM}}
\label{quarkvertex}
\end{figure}

\noindent
If we now integrate out the charged zero modes, we get a connected diagram and taking the field theory limit (where $\alpha' \rightarrow 0)$ we receive an MSSM process which contributes to neutral Kaon mixing (figure \ref{quarkvertex}).

Since we are  interested in the order of magnitude of the process,
we just estimate  the characteristic contributions to the vertices and the
propagators. The instanton suppression factors of the nonperturbative
Yukawa-coupling and the $\mu$-term are fixed by the requirement to have the
order of magnitude of the respective standard model/MSSM values. By this
reasoning we obtain
\begin{equation}
\mathcal{M} \approx \frac{e^{-S_0}}{M_{F}}
\frac{\mu ~ m_{\textrm{susy}} ^4 m_{K^0}^2 }{m_{\tilde{q} _L ^1} ^2
  m_{\tilde{u}_R ^{12}} ^2 m_{\tilde{H}}^2} \frac{ m_d
  m_s}{m_{\textrm{weak}}^2}\; ,
\end{equation} 
where we wrote a factor of $m_x ^2$ for a propagator of a boson and $m_x$ for
the propagator of a fermion. Moreover, we inserted a factor of the
characteristic energy $m_{\textrm{susy}} ^4$ coming from the loop
integration. In addition there is a factor of the quark mass over the
electroweak breaking scale coming from the respective Yukawa coupling and
$\mu$ from the Higgs $\mu$-term. 
$M_{F}$ is the  mass suppression of the physical dimension five operator.
Finally we need a factor of $g_s ^4$ (where $g_s$ is the string coupling) because we inserted 8 charged zero modes on an annulus, which comes with an overall normalization of $g_s ^0$. The factor of $e^{-S_0}$ can also be expressed in terms of quark masses $m_c : m_t \approx \frac{1}{100}$ which is required to get the right family mass hierarchy.
\vskip 10 pt 
We are now going to estimate the contribution of this process to neutral Kaon mixing which can be seen experimentally in the mass difference $\Delta m$ of the $K_L ^0$ and $K_S ^0$ states. It is well known how to compute the mass difference from the effective 4-fermion operator of $K^0 \leftrightarrow \bar{K} ^0$-mixing: The evaluation of the operator between the $K^0$ eigenstates gives a factor of $\frac{8}{3} \frac{f_K ^2 m_K ^2}{2 m_K}$ ($f_K$ is the Kaon decay constant and $m_K$ is the Kaon mass) and together with the estimate of the amplitude this gives 
\begin{equation}
\Delta m \approx \frac{e^{-S_0}}{M_F}    \frac{\mu ~ m_{\textrm{susy}} ^4  }{m_{\tilde{q} _L ^1} ^2 m_{\tilde{u}_R ^{12}} ^2 m_{\tilde{H}}^2} \frac{ m_d m_s}{m_{\textrm{weak}}^2} \frac{8}{3} \frac{f_K ^2 m_{K^0} ^2}{2 m_K}
\end{equation}
This gives a contribution to the mass difference depending on the string mass of the order
\begin{equation}
\Delta m \approx \frac{1}{M_F} 10^{-21}\, \textrm{TeV}^{2}\; .
\end{equation}
Now the experimental accuracy of measurements of the mass difference is of
the order $10^{-21}$ TeV which for $M_F=M_w$
leads to a lower bound on the string scale of 
\begin{equation}
M_s \geq  10^{-7}\, \textrm{TeV}.
\end{equation}
This is not very restrictive. The first reason for this is the high suppression
of the three-instanton amplitude due to the down- and strange- quark Yukawa
couplings. In addition the winding scale factor $M_w$ gives a 
suppression of the order
of $M_P ^{1/3}$. Contrarily, for $M_F=M_s$ 
we  get a lower bound on the string scale of the order of
\begin{equation}
M_s \geq 1 \, \textrm{TeV},
\end{equation}
which is significantly larger.
Finally the result also depends on the experimental accuracy of $\Delta m $ measurements and can give stronger bounds in the future.

\subsection{Lepton flavour violation}
In the standard model with massless neutrinos, we have conserved lepton flavour number. Thus every observation of lepton-flavour violating (LFV) processes would be a definite hint for physics beyond the standard model. The most precise bounds on possible LFV processes come from the Kaon and D-meson systems. In the following we list the most promising candidate decay processes (the branching ratio BR gives the ratio of the decay rate of the process to the total decay rate of the initial meson). 
\begin{table}[h]
\centering
\caption{Kaon decays into standard model particles with BR $< 5\cdot10^{-10}$ \cite{Barker:2000gd}}
\vskip0.3cm
\label{k0}
\begin{tabular}{|c|c|}
\hline
Decay & Branching ratio \\
\hline
$K_L ^0 \rightarrow \mu ^{\pm} e^{\mp} $ & $ < 4.7 \cdot 10 ^{-12} $\\
$K^+ \rightarrow \pi ^+ \mu ^+ e ^- $ & $ < 2.8 \cdot 10 ^{-11}  $\\
$ K_L ^0 \rightarrow \pi ^0 \mu ^+ e^- $ & $ <4.4 \cdot 10^{-10} $ \\
$K^+\rightarrow \pi ^- \mu ^+ e^+$ & $ < 5.0 \cdot 10^{-10} $ \\
\hline
\end{tabular}
\end{table}

\begin{table}[h]
\centering
\label{d0}
\caption{Examples for decay modes of $D^0$ with BR$<5\cdot 10^{-6}$ \cite{Burdman:2003rs}}
\vskip0.3cm
\begin{tabular}{|c|c|}
\hline
Decay & Branching ratio \\
\hline
$D^0 \rightarrow e ^+ e^- $&$ <0.6 \cdot 10 ^{-6}$ \\
$D^0 \rightarrow \mu ^+ \mu ^- $&$< 3.4 \cdot 10 ^{-6}$ \\
$D^0 \rightarrow \mu ^{\pm} e^{\mp} $&$ <1.9 \cdot 10 ^{-6}$ \\
\hline
\end{tabular}
\end{table}

\begin{figure}[h]
\centering
	\includegraphics[width=0.90\textwidth]{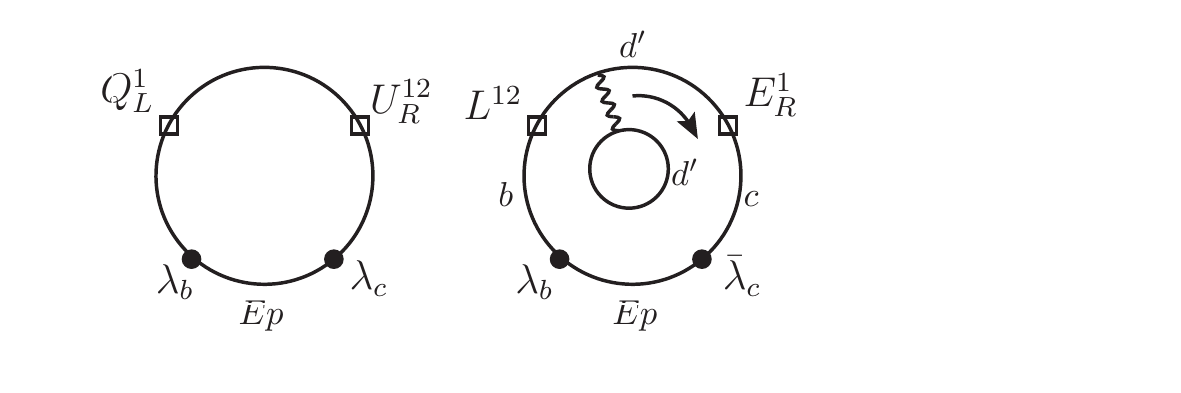}
	\caption{Instanton contribution to $D^0$-decay}
\label{annulusneu}

\end{figure}

\begin{figure}[h]
\centering

		\includegraphics[width=0.70\textwidth]{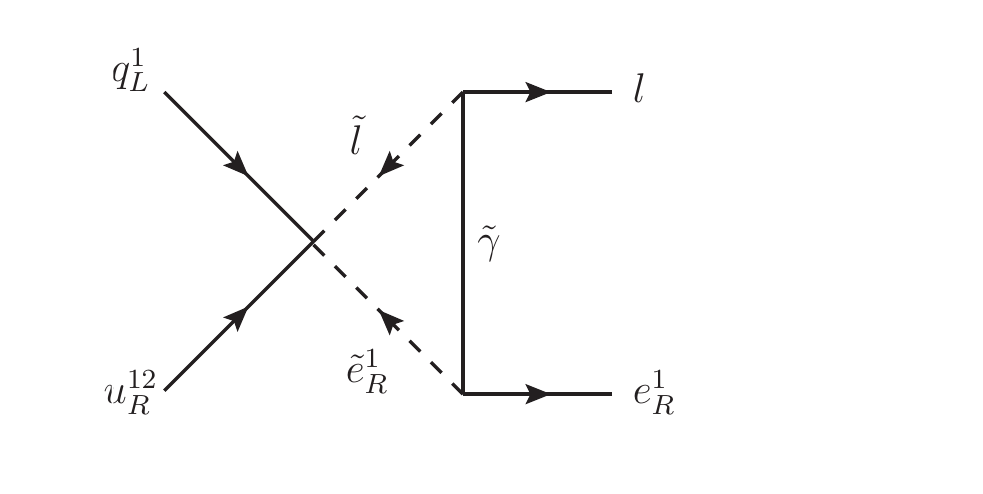}
	\caption{Instanton contribution to $D^0$-decay in the field theory limit}
\label{zerfall}
\end{figure}

The list of higher dimensional operators from section 3.2 reveals the existence of various interesting operators, which contain two quark and two lepton fields and therefore could contribute to one of the meson decays of the table.
%In the last section, we observed that there is an additional limitation on possible processes if we want to realize them purely string theoretic, and therefore we will directly look for combinations of discs or discs with annuli. 
Taking the operator 
\begin{equation}
Q_L ^1 U_R ^{12} E_R ^1 L^{12}
\end{equation}
we again get a specific combination of a disc with an annulus which is given in figure \ref{annulusneu}.
There are the following particles running in the loop (we use our example quiver of table \ref{5stack}): The $bd'$-sector corresponds to a lepton $L$, $dd'$ is a gauge boson (photon) or its superpartner and the $dc$-sector corresponds to $E_R ^1$. Therefore, in field theory the process would correspond to the Feynman diagram given in figure \ref{zerfall}. To get proper MSSM couplings with standard model fermions as out-states, we get superpartners running in the loop: In our case sleptons $\tilde{L}$ and $\tilde{E}_R ^1$ and the photino $\tilde{\gamma}$. The whole process would therefore contribute to the following decays of the $D^0$-meson:
\begin{equation}
D^0 \rightarrow e^- e^+ \, ,\quad D^0 \rightarrow \mu ^- e^+\, .
\end{equation}

\vskip 20pt

Let us now estimate the order of magnitude of this process in field theory, as given in figure \ref{zerfall}. From the instanton vertex we get the suppression factor $\frac{e^{-S_0}}{M_F}$ . From the loop integration we get a factor of $m_{\textrm{susy}} ^4 $. Together with the propagator masses, the electromagnetic coupling constant and using as kinematical factor the mass of the initial state of the $D^0$ we get
\begin{equation}
\mathcal{M} \propto \frac{e^{-S_0}}{M_F} m_{\textrm{susy}}^4 \frac{m_{D^0} ^2}{m_{\textrm{susy} }^5} g_{\textrm{em}} ^2 (m_{\textrm{susy}})
\end{equation}
Using $1 \textrm{TeV}$ for the supersymmetry scale, $\frac{1}{100}$ for the instantonic suppression (which is set by the ratio of the heaviest quark to the charm quark)  and the mass of the $D^0$ (2GeV), we get as a rough estimate
\begin{equation}
\mathcal{M} \propto 4\cdot 10 ^{-9} \frac{\textrm{TeV}}{M_F}
\end{equation}
We want to compare this to the experimental bound for the above given $D^0$-decays. The lifetime of the $D^0$ is $\tau_{D ^0} = 4,1 \cdot 10^{-13} sec$, which leads to (using the branching ratio given in table 6.3) a decay rate of 
\begin{equation}
\Gamma_{D^0 \rightarrow  LE_R ^1 } = \textrm{BR}(D^0 \rightarrow L E_R ^1) \Gamma _{D^0} <  10^{-18} \textrm{GeV}.
\end{equation}
Now the amplitude of the process can be estimated by using
\begin{equation}
\Gamma_{D^0 \rightarrow L E_R ^1 } \approx m_{D^0} |\mathcal{M}|^2
\end{equation}
The result is $|\mathcal{M}| <  10^{-9}$.
Comparing this with the field theory estimate of the instanton process (and assuming $M_F = M_w$) we get a lower bound on the string scale of order 
\begin{equation}
M_s \geq 10^{-7} \textrm{TeV}
\end{equation}
 which is certainly not very restrictive.

We see that the winding factor gives us an additional very high suppression factor such that instanton contributions to meson decay are very weak. To illustrate this, we skip this  and calculate only with the naive expectation of a suppression of $M_F = M_s$ characteristic for effective dimension 5 operators, i.e. we have
\begin{equation}
\mathcal{M} \propto \frac{e^{-S_0}}{M_s} m_{\textrm{susy}}^4 \frac{m_{D^0} ^2}{m_{\textrm{susy} }^5} g_{\textrm{em}} ^2 (m_{\textrm{susy}}) = 4\cdot 10^{-9} \frac{\textrm{TeV}}{M_s}\, .
\end{equation}
This would lead to the lower bound of 
\begin{equation}
M_s \geq 4 \textrm{TeV}.
\end{equation}
Another important fact is that the experimental bounds on the $D^0$-decay are not as good as for Kaon decays. If we possibly had  future experimental bounds similar to Kaon decays (table \ref{k0}), we would get $M_s >3\cdot 10^{-3}$ TeV with winding scale and $M_s > 4\cdot 10^3$ TeV without this scale. 

\begin{figure}[h]
\centering
		\includegraphics[width=0.90\textwidth]{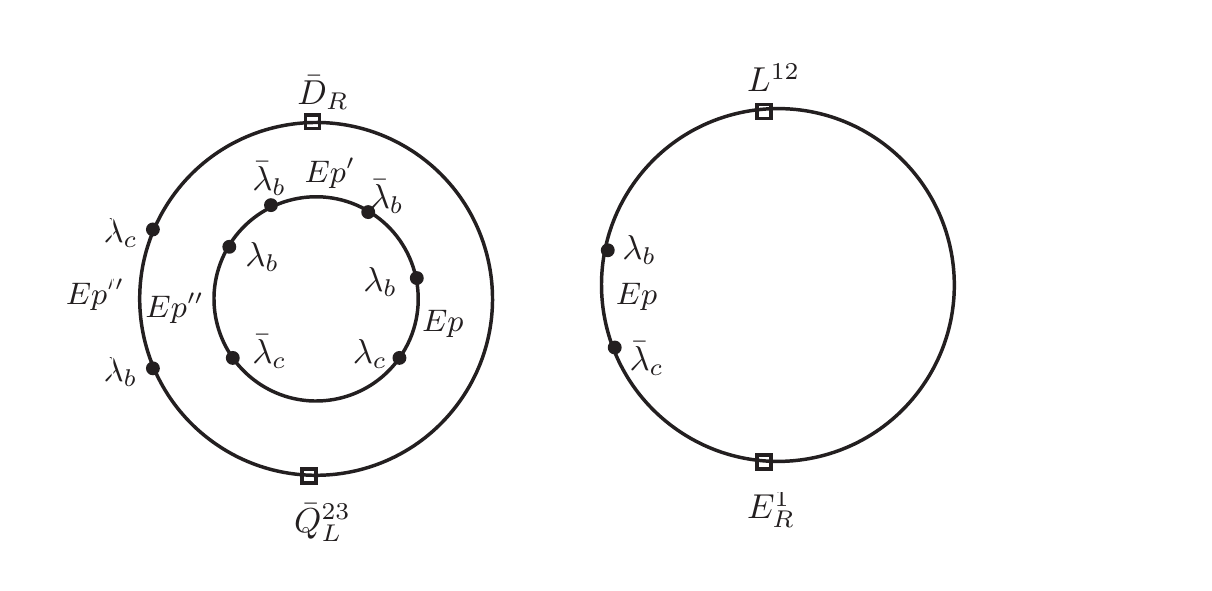}
	
\caption{\small{3-instanton contribution to the LFV Kaon decay}}
\label{k0zerfall}

\end{figure}

\begin{figure}[h]
\centering
		\includegraphics[width=0.70\textwidth]{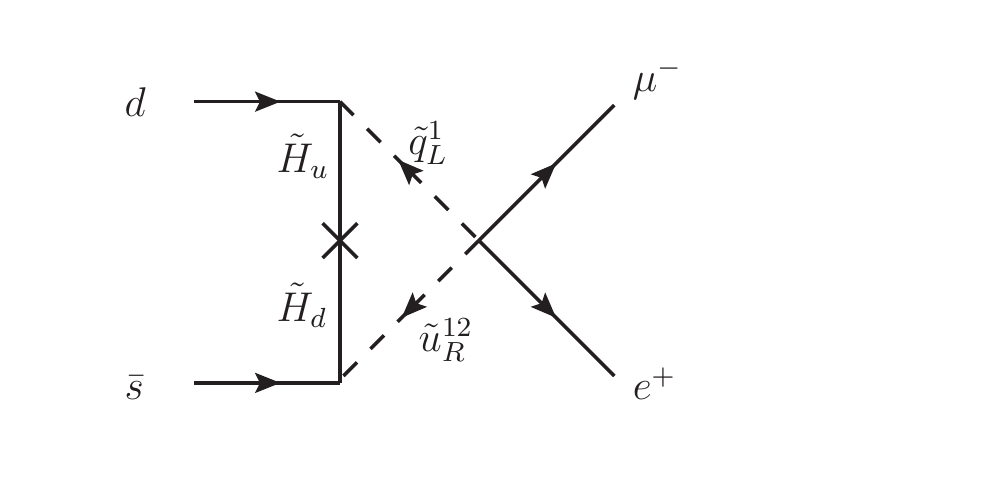}
	\caption{Instanton contribution to $K^0 \rightarrow e^+ \mu ^-$}
\label{k0zerfall2}
\end{figure}

A process which is restricted much more by experiment is the LFV Kaon decay
\begin{equation}
K^0 \rightarrow e^+ \mu^-
\end{equation}
We propose that this process gets contributions from a 3-instanton process. To show this, we take the operator $Q_L ^1 U_R ^{12} E_R ^1 L^{12}$ found in eq.\eqref{ph}. Similar to the contribution to neutral Kaon mixing discussed in the last section we combine a disc and an annulus diagram (figure \ref{k0zerfall}). Two of the zero modes which are needed to generate the operator are inserted in the disc and the other two in the annulus, and therefore we get again a connected diagram after the integration over the zero modes. In the field theory limit this corresponds to the process given in figure \ref{k0zerfall2}.

Let us also determine the order of magnitude of this process. If we use the same estimates as for neutral Kaon mixing and $D^0$-decay we get the following for the amplitude:

\begin{equation}
\mathcal{M} \propto \frac{e^{-S_0}}{M_F}\frac{\mu ~m_{\textrm{susy}} ^4 m_{K^0} ^2 }{m_{\textrm{susy}}^4 m_{\textrm{susy}}^2} \frac{ m_d m_s}{m_{weak} ^2}
\end{equation}

In addition to the estimate for $D^0$ we use here the $\mu$-term which we assume to be of the order of the supersymmetry scale and we used the standard model values for the Yukawa couplings which are of the order of $\frac{ m_{\textrm{quark}}}{m_{\textrm{weak}}}$, where $m_{\textrm{weak}}$ is the electroweak breaking scale.

Taking again the same estimate for the instanton suppression factor we get for the estimate of the amplitude $\mathcal{M} \propto 10^{-12}  $ and a bound on the string scale of the order
\begin{equation}
M_s \geq 10^{-12} \textrm{TeV}
\end{equation}
This is clearly not a bound on the string scale at all. Again the extreme suppression of the amplitude can be explained by the two light Yukawa couplings included in the 3-instanton process and the contribution of the winding scale factor.

In the following table \ref{bound} we compare the perturbative bound \cite{Abel:2003fk} with the instanton results from $D^0$-decay (first line with the real experimental upper bound, and second line with a hypothetical bound comparable to Kaon decay) and with the results for LFV Kaon decays (third line) and neutral Kaon mixing (fourth line). These could become important in MSSM models, for which there is no tree level perturbative contribution, and hence  instanton contributions to lepton flavour violating Kaon decays are dominant. To emphasize the importance of the winding factor we list the results with and without this factor.

\begin{table}[h]
\centering
\caption{Comparison of the results for the string scale}
\label{bound}
\vskip0.3cm
\begin{tabular}{|c|c|c|c|}
\hline
Process & Perturbative & Winding& No winding \\
\hline
$D^0 \rightarrow L E_R ^1$ & - &$ 10^{-7}$ TeV & $4$ TeV \\
$D^0 \rightarrow L E_R ^1$ &-& $ 10^{-3}$ TeV & $4 \cdot 10^3$ TeV \\
$K_L ^0 \rightarrow \mu ^- e^+$ & $100$ TeV& $ 10^{-12}$ TeV & $ 10 ^{-2} $ TeV \\
$K^0 \leftrightarrow \bar{K} ^0 $ & $10^3$ TeV& $ 10^{-7}$ TeV & $1 $TeV\\ 
\hline
\end{tabular}
\end{table}
 
We conclude that the results using $D^0,K^0$-decay and neutral Kaon mixing with today's experimental bounds are not very restrictive, especially in comparison to results from perturbative allowed operators. However, for models which do not allow such operators and also would give rise to non-perturbative Kaon or D-meson decay operators the above results would give restrictions on the string mass.

\section{Conclusion and outlook}
In this paper we have considered instanton induced FCNC couplings
in (softly broken) supersymmetric intersecting D-brane models. 
Technically, the relevant field theory diagrams involve dimension
five F-terms dressed by an additional loop to convert the
bosonic squarks and sleptons to fermionic matter particles.
We have seen that in string theory the corresponding diagrams
can be consistently described in an extended multi-instanton
calculus, which involved disc and annulus diagrams with
a mixture of space-time filling and instantonic D-brane boundaries
with the appropriate insertions  of (boundary changing) matter
and instanton zero modes. We did not provide the full formalism
but discussed this for  a couple of relevant examples.
 
Equipped with these techniques we considered a concrete five
stack MSSM-like intersecting D-brane model and found
that experimentally highly suppressed FCNC processes like
$K^0-\ov K^0$ mixing and leptonic meson decays are generated
by (multi-) D-brane instantons. The appearing exponential
suppression factors could be estimated due to  the fact that
the corresponding instantons also induced perturbatively
forbidden Yukawa couplings. We computed and compared the appearing bounds
on the string scales for two reasonable mass scales of the
physical dimension five operators. For  this latter scale
being of the order of the string scale, one finds
bounds of the order $1-4\, $TeV, which are of similar size as the
bounds for perturbative dimension six operators. 
However, for quiver type MSSMs  the suppression scale
is rather the significantly larger winding scale, which only give 
extremely  mild lower bounds on the string scale. 
%We observed that instanton contributions to neutral kaon mixing, $D^0$ and $K^0$-decay give very weak bounds
%on the string scale. In general this is due to the winding scale supression of holomorphic
%amplitudes and the present experimental bounds. 
Moreover, for processes involving more than one instanton, like neutral Kaon
mixing and the LVF Kaon decay we get additional high suppression due to the
light quark Yukawa couplings generated by D-instantons.  

%In particular the non-perturbative contributions are lower than those coming
%from perturbatively allowed operators \cite{Abel:2003fk}. 

We conclude that, if there exist MSSM models which avoid perturbative 
allowed FCNC operators such as $Q_L \bar{Q}_L Q_L
\bar{Q}_L$ or $D_R \bar{D}_R D_R \bar{D}_R$ (which would be very attractive
for a string model in the range of the LHC), instanton contributions  
can become important and, depending on the mass scale of the dimension five operators, can provide non-negligible lower bounds on the string scale. 
%Such models could be realized for example, if
%one assigns different realizations to different families.
% such as
%bifundamental representations $(a,b)$ for the first family $Q_L ^{1}$ and
%$(a,\bar{b})$ for the other families $Q_L ^{2,3}$ and similar 
%for $D_R$. 

%%%%%%%%%%%%%%%%%%%%%%%%%%%%%%%%%%%%%%%%%%%%%%%
\subsection*{Acknowledgement}
%%%%%%%%%%%%%%%%%%%%%%%%%%%%%%%%%%%%%%%%%%%%%%%
R.~Blumenhagen and D.~L\"ust  would like to thank the Kavli Institute for
Theoretical Physics, Santa Barbara for the hospitality during part of this work. This research was supported in part by the National Science Foundation under Grant No. PHY05-51164.


\begin{thebibliography}{99}






%\cite{Blumenhagen:2006ci}
\bibitem{Blumenhagen:2006ci}
  R.~Blumenhagen, B.~K\"ors, D.~L\"ust and S.~Stieberger,
  ``Four-dimensional String Compactifications with D-Branes, Orientifolds   and
  Fluxes,''
  Phys.\ Rept.\  {\bf 445}, 1 (2007)
  [arXiv:hep-th/0610327].
  %%CITATION = PRPLC,445,1;%%
  
  %\cite{Lust:2008qc}
\bibitem{Lust:2008qc}
  D.~L\"ust, S.~Stieberger and T.~R.~Taylor,
  ``The LHC String Hunter's Companion,''
  Nucl.\ Phys.\  B {\bf 808}, 1 (2009)
  [arXiv:0807.3333 [hep-th]].
  %%CITATION = NUPHA,B808,1;%%
  
  %\cite{Lust:2009pz}
\bibitem{Lust:2009pz}
  D.~L\"ust, O.~Schlotterer, S.~Stieberger and T.~R.~Taylor,
  ``The LHC String Hunter's Companion (II): Five-Particle Amplitudes and
  Universal Properties,''
  Nucl.\ Phys.\  B {\bf 828}, 139 (2010)
  [arXiv:0908.0409 [hep-th]].
  %%CITATION = NUPHA,B828,139;%%
  
  
  %\cite{Anchordoqui:2008hi}
\bibitem{Anchordoqui:2008hi}
  L.~A.~Anchordoqui, H.~Goldberg and T.~R.~Taylor,
  ``Decay widths of lowest massive Regge excitations of open strings,''
  Phys.\ Lett.\  B {\bf 668}, 373 (2008)
  [arXiv:0806.3420 [hep-ph]].
  %%CITATION = PHLTA,B668,373;%%
  
  
%\cite{Anchordoqui:2008di}
\bibitem{Anchordoqui:2008di}
  L.~A.~Anchordoqui, H.~Goldberg, D.~L\"ust, S.~Nawata, S.~Stieberger and T.~R.~Taylor,
  ``Dijet signals for low mass strings at the LHC,''
  Phys.\ Rev.\ Lett.\  {\bf 101}, 241803 (2008)
  [arXiv:0808.0497 [hep-ph]].
  %%CITATION = PRLTA,101,241803;%%
  
  
  %\cite{Anchordoqui:2009mm}
\bibitem{Anchordoqui:2009mm}
  L.~A.~Anchordoqui, H.~Goldberg, D.~L\"ust, S.~Nawata, S.~Stieberger and T.~R.~Taylor,
  ``LHC Phenomenology for String Hunters,''
  Nucl.\ Phys.\  B {\bf 821}, 181 (2009)
  [arXiv:0904.3547 [hep-ph]].
  %%CITATION = NUPHA,B821,181;%%
  
  
  
  %\cite{Cullen:2000ef}
\bibitem{Cullen:2000ef}
  S.~Cullen, M.~Perelstein and M.~E.~Peskin,
  ``TeV strings and collider probes of large extra dimensions,''
  Phys.\ Rev.\  D {\bf 62}, 055012 (2000)
  [arXiv:hep-ph/0001166].
  %%CITATION = PHRVA,D62,055012;%%
  
 

%\cite{Abel:2003fk}
\bibitem{Abel:2003fk}
  S.~A.~Abel, M.~Masip and J.~Santiago,
  ``Flavour changing neutral currents in intersecting brane models,''
  JHEP {\bf 0304}, 057 (2003)
  [arXiv:hep-ph/0303087].
  %%CITATION = JHEPA,0304,057;%%
  
  







%\cite{Blumenhagen:2006xt}
\bibitem{Blumenhagen:2006xt}
  R.~Blumenhagen, M.~Cvetic and T.~Weigand,
  ``Spacetime instanton corrections in 4D string vacua - the seesaw mechanism
  for D-brane models,''
  Nucl.\ Phys.\  B {\bf 771}, 113 (2007)
  [arXiv:hep-th/0609191].
  %%CITATION = NUPHA,B771,113;%%
  
  
  %\cite{Ibanez:2006da}
\bibitem{Ibanez:2006da}
  L.~E.~Ibanez and A.~M.~Uranga,
  ``Neutrino Majorana masses from string theory instanton effects,''
  JHEP {\bf 0703}, 052 (2007)
  [arXiv:hep-th/0609213].
  %%CITATION = JHEPA,0703,052;%%
  
  
  %\cite{Blumenhagen:2007zk}
\bibitem{Blumenhagen:2007zk}
  R.~Blumenhagen, M.~Cvetic, D.~L\"ust, R.~Richter and T.~Weigand,
  ``Non-perturbative Yukawa Couplings from String Instantons,''
  Phys.\ Rev.\ Lett.\  {\bf 100}, 061602 (2008)
  [arXiv:0707.1871 [hep-th]].
  %%CITATION = PRLTA,100,061602;%%

%\cite{Florea:2006si}
\bibitem{Florea:2006si}
  B.~Florea, S.~Kachru, J.~McGreevy and N.~Saulina,
  ``Stringy Instantons and Quiver Gauge Theories,''
  JHEP {\bf 0705} (2007) 024
  [arXiv:hep-th/0610003].
  %%CITATION = JHEPA,0705,024;%%

%\cite{Blumenhagen:2009qh}
\bibitem{Blumenhagen:2009qh}
  R.~Blumenhagen, M.~Cvetic, S.~Kachru and T.~Weigand,
  ``{\small D}-Brane Instantons in Type {II} Orientifolds,''
  Ann.\ Rev.\ Nucl.\ Part.\ Sci.\  {\bf 59} (2009) 269
  [arXiv:0902.3251 [hep-th]].
  %%CITATION = ARNUA,59,269;%%


  %\cite{Anastasopoulos:2009mr}
\bibitem{Anastasopoulos:2009mr}
  P.~Anastasopoulos, E.~Kiritsis and A.~Lionetto,
  ``On mass hierarchies in orientifold vacua,''
  JHEP {\bf 0908} (2009) 026
  [arXiv:0905.3044 [hep-th]].
  %%CITATION = JHEPA,0908,026;%%
  
  %\cite{Cvetic:2009yh}
\bibitem{Cvetic:2009yh}
  M.~Cvetic, J.~Halverson and R.~Richter,
  ``Realistic Yukawa structures from orientifold compactifications,''
  JHEP {\bf 0912} (2009) 063
  [arXiv:0905.3379 [hep-th]].
  %%CITATION = JHEPA,0912,063;%%

%\cite{Blumenhagen:2007bn}
\bibitem{Blumenhagen:2007bn}
  R.~Blumenhagen, M.~Cvetic, R.~Richter and T.~Weigand,
  ``Lifting D-Instanton Zero Modes by Recombination and Background Fluxes,''
  JHEP {\bf 0710}, 098 (2007)
  [arXiv:0708.0403 [hep-th]].
  %%CITATION = JHEPA,0710,098;%%

%\cite{GarciaEtxebarria:2007zv}
\bibitem{GarciaEtxebarria:2007zv}
  I.~Garcia-Etxebarria and A.~M.~Uranga,
  ``Non-perturbative superpotentials across lines of marginal stability,''
  JHEP {\bf 0801}, 033 (2008)
  [arXiv:0711.1430 [hep-th]].
  %%CITATION = JHEPA,0801,033;%%



%\cite{Conlon:2006tj}
\bibitem{Conlon:2006tj}
  J.~P.~Conlon, D.~Cremades and F.~Quevedo,
  ``Kaehler potentials of chiral matter fields for Calabi-Yau string
  compactifications,''
  JHEP {\bf 0701} (2007) 022
  [arXiv:hep-th/0609180].
  %%CITATION = JHEPA,0701,022;%%

%\cite{Conlon:2007zza}
\bibitem{Conlon:2007zza}
  J.~P.~Conlon and D.~Cremades,
  ``The neutrino suppression scale from large volumes,''
  Phys.\ Rev.\ Lett.\  {\bf 99} (2007) 041803
  [arXiv:hep-ph/0611144].
  %%CITATION = PRLTA,99,041803;%%



%\cite{Conlon:2009qa}
\bibitem{Conlon:2009qa}
  J.~P.~Conlon and E.~Palti,
  ``On Gauge Threshold Corrections for Local IIB/F-theory GUTs,''
  Phys.\ Rev.\  D {\bf 80}, 106004 (2009)
  [arXiv:0907.1362 [hep-th]].
  %%CITATION = PHRVA,D80,106004;%%


%\cite{Conlon:2010ji}
\bibitem{Conlon:2010ji}
  J.~P.~Conlon and F.~G.~Pedro,
  ``Moduli Redefinitions and Moduli Stabilisation,''
  arXiv:1003.0388 [hep-th].
  %%CITATION = ARXIV:1003.0388;%%
  
   %\cite{Billo:2002hm}
\bibitem{Billo:2002hm}
  M.~Billo, M.~Frau, I.~Pesando, F.~Fucito, A.~Lerda and A.~Liccardo,
  ``Classical gauge instantons from open strings,''
  JHEP {\bf 0302} (2003) 045
  [arXiv:hep-th/0211250].
  %%CITATION = JHEPA,0302,045;%%

  
  
  




%\cite{Argurio:2007vqa}
\bibitem{Argurio:2007vqa}
  R.~Argurio, M.~Bertolini, G.~Ferretti, A.~Lerda and C.~Petersson,
  ``Stringy Instantons at Orbifold Singularities,''
  JHEP {\bf 0706}, 067 (2007)
  [arXiv:0704.0262 [hep-th]].
  %%CITATION = JHEPA,0706,067;%%

%\cite{Bianchi:2007wy}
\bibitem{Bianchi:2007wy}
  M.~Bianchi, F.~Fucito and J.~F.~Morales,
  ``D-brane Instantons on the $T^6/Z_3$ orientifold,''
  JHEP {\bf 0707}, 038 (2007)
  [arXiv:0704.0784 [hep-th]].
  %%CITATION = JHEPA,0707,038;%%


%\cite{Ibanez:2007rs}
\bibitem{Ibanez:2007rs}
  L.~E.~Ibanez, A.~N.~Schellekens and A.~M.~Uranga,
  ``Instanton Induced Neutrino Majorana Masses in CFT Orientifolds with
  MSSM-like spectra,''
  JHEP {\bf 0706}, 011 (2007)
  [arXiv:0704.1079 [hep-th]].
  %%CITATION = JHEPA,0706,011;%%

%\cite{Akerblom:2007uc}
\bibitem{Akerblom:2007uc}
  N.~Akerblom, R.~Blumenhagen, D.~Lust and M.~Schmidt-Sommerfeld,
  ``Instantons and Holomorphic Couplings in Intersecting D-brane Models,''
  JHEP {\bf 0708}, 044 (2007)
  [arXiv:0705.2366 [hep-th]].
  %%CITATION = JHEPA,0708,044;%%


  
%\cite{Cvetic:2010mm}
\bibitem{Cvetic:2010mm}
  M.~Cvetic, J.~Halverson, P.~Langacker and R.~Richter,
  ``The Weinberg Operator and a Lower String Scale in Orientifold
  Compactifications,''
  arXiv:1001.3148 [hep-th].
  %%CITATION = ARXIV:1001.3148;%%


%\bibitem{feynmanrules}
%J.~Rosiek "Complete set of Feynman rules for the MSSM - \emph{erratum}" \newline
%[arXiv: hep-ph/9511250v3] \newline
%M.~Kuroda "Complete Lagrangian of MSSM"
%[arXiv: hep-ph/9902340v3]



  
  %\cite{Barker:2000gd}
\bibitem{Barker:2000gd}
  A.~R.~Barker and S.~H.~Kettell,
  ``Developments in Rare Kaon Decay Physics,''
  Ann.\ Rev.\ Nucl.\ Part.\ Sci.\  {\bf 50} (2000) 249
  [arXiv:hep-ex/0009024].
  %%CITATION = ARNUA,50,249;%%
  
  %\cite{Burdman:2003rs}
\bibitem{Burdman:2003rs}
  G.~Burdman and I.~Shipsey,
  ``$D^0$ - $\bar{D}^0$ mixing and rare charm decays,''
  Ann.\ Rev.\ Nucl.\ Part.\ Sci.\  {\bf 53} (2003) 431
  [arXiv:hep-ph/0310076].
  %%CITATION = ARNUA,53,431;%%

\end{thebibliography}
\end{document}